\documentclass[12pt,english]{article}
\pdfoutput=1
\usepackage[T1]{fontenc}
\usepackage{fullpage}
\usepackage{authblk}
\usepackage{graphicx,array}
\usepackage{cite}
\usepackage{color}
\usepackage{caption}
\usepackage{latexsym}
\usepackage{amsthm}
\usepackage{amsmath}
\usepackage[titletoc]{appendix}
\usepackage{enumitem}
\usepackage{amssymb}
\usepackage{color}
\usepackage[unicode=true,
 bookmarks=true,bookmarksnumbered=false,bookmarksopen=false,
 breaklinks=false,pdfborder={0 0 1},backref=false,colorlinks=true]
 {hyperref}
\hypersetup{pdftitle={Towards a Bit Threads Derivation of Holographic Entanglement of Purification},
 pdfauthor={Ning Bao, Aidan Chatwin-Davies, Jason Pollack, Grant N. Remmen},
 citecolor=black,linkcolor=black,urlcolor=black}
\usepackage{breakurl}
\usepackage[hang,flushmargin]{footmisc} 
\usepackage{listings}
\usepackage{mathrsfs}
\usepackage{amsthm}
\usepackage{afterpage}

\newcommand{\Eq}[1]{Eq.~(\ref{#1})}

\newcommand{\Sec}[1]{Sec.~\ref{#1}}

\newcommand{\Fig}[1]{Fig.~\ref{#1}}
\newcommand{\App}[1]{App.~\ref{#1}}
\newcommand{\Ref}[1]{Ref.~\cite{#1}}
\newcommand{\Refs}[1]{Refs.~\cite{#1}}
\usepackage{accents}

\newcommand{\bra}[1]{\langle #1 |}
\newcommand{\ket}[1]{| #1 \rangle}

\newcommand{\ketbra}[2]{|#1\rangle \langle #2 |}

\newtheorem{thm}{Theorem}[section]

\newtheorem{example}[thm]{Example}

\newcommand{\pr}{\prime}

\newcommand{\mrm}[1]{\mathrm{#1}}
\newcommand{\Hil}{\mathcal{H}}
\usepackage{accents}

\newcommand{\be}{\begin{equation}}
\newcommand{\ee}{\end{equation}}

\DeclareMathOperator{\Tr}{Tr}

\usepackage{xcolor}
\usepackage{color}

\begin{document}

\interfootnotelinepenalty=10000
\hfill

\vspace{2cm}
\thispagestyle{empty}
\begin{center}
{\LARGE \bf
Towards a Bit Threads Derivation of \\[2mm] Holographic Entanglement of Purification
}\\
\bigskip\vspace{1cm}{
{\large Ning Bao,${}^{a,b}$ Aidan Chatwin-Davies,${}^c$ Jason Pollack,${}^d$ and Grant N. Remmen${}^a$}
} \\[7mm]
 {\it ${}^a$Center for Theoretical Physics and Department of Physics \\
     University of California, Berkeley, CA 94720, USA and \\
     Lawrence Berkeley National Laboratory, Berkeley, CA 94720, USA \\[1.5 mm]
 ${}^b$Computational Science Initiative, Brookhaven National Lab, Upton, NY 11973, USA
 \\[1.5mm] ${}^c$KU Leuven, Institute for Theoretical Physics\\Celestijnenlaan 200D B-3001 Leuven, Belgium\\[1.5mm]
 ${}^d$Department of Physics and Astronomy\\University of British Columbia, Vancouver, BC V6T 1Z1, Canada} \let\thefootnote\relax\footnote{\noindent e-mail: \url{ningbao75@gmail.com}, \url{aidan.chatwindavies@kuleuven.be}, \url{jpollack@phas.ubc.ca}, \\ \hphantom{e-mail:} \url{grant.remmen@berkeley.edu}} \\
 \end{center}
\bigskip
\centerline{\large\bf Abstract}
\begin{quote} \small
We apply the bit thread formulation of holographic entanglement entropy to reduced states describing only the geometry contained within an entanglement wedge.
We argue that a certain optimized bit thread configuration, which we construct, gives a purification of the reduced state to a full holographic state obeying a precise set of conditional mutual information relations.
When this purification exists, we establish, under certain assumptions, the conjectured $E_P = E_W$ relation equating the entanglement of purification with the area of the minimal cross section partitioning the bulk entanglement wedge.
Along the way, we comment on minimal purifications of holographic states, geometric purifications, and black hole geometries.

\end{quote}
	
\setcounter{footnote}{0}

\newpage
\tableofcontents
\newpage
\baselineskip=18pt
\section{Introduction}
The relation between entanglement and (holographic) spacetime \cite{tHooft:1993dmi,Susskind:1994vu,Bousso:2002ju} is a powerful connection that has been extended through many interesting developments over the past decade.
Perhaps the most elegant of these extensions is the Ryu-Takayanagi (RT) formula, 
\be 
S[\rho]=\frac{\cal A}{4G_N\hbar},
\label{eq:RT}
\ee
relating the entanglement entropy $S[\rho]=-\Tr \rho \log \rho$ of the reduced density matrix $\rho$ of a holographic large-$N$ CFT in a boundary region to the area ${\cal A}$ of the minimal bulk surface homologous to that region~\cite{Ryu:2006bv,Ryu:2006ef,Lewkowycz:2013nqa}.
This result, among others, has led to a concrete instantiation of the emergence of spacetime from entanglement in the context of AdS/CFT~\cite{Maldacena:1997re,Gubser:1998bc,Witten:1998qj,Aharony:1999ti}.

Recently, the connection between boundary information-theoretic quantities and bulk areas has been extended to a conjectured ``$E_P = E_W$'' relationship between the area of bulk-anchored minimal surfaces and the entanglement of purification \cite{terhal2002entanglement}, as in Refs.~\cite{Takayanagi:2017knl, nguyen2018entanglement}. The entanglement of purification is an entanglement measure that characterizes the degree of entanglement between two subsystems of a generically mixed state. It is particularly useful when only partial information about a state is known. As we review below, the entanglement of purification is a bound on the entanglement between subsystems in a given purification; purifications that saturate this bound comprise a class of optimal purifications or completions of the state in question.
The $E_P = E_W$ conjecture has been further extended, studied, and generalized in recent work~\cite{Agon:2018lwq, bao2018holographic, bao2018conditional, bao2019entanglement, umemoto2018entanglement,Faulkner,Kudler-Flam:2019oru,Du:2019emy,Jokela:2019ebz}.

In this paper, we will seek to make progress towards proving the $E_P = E_W$ conjecture using technology from the bit threads program as described in \Ref{Freedman:2016zud}. This program, which in its original formalism~\cite{Freedman:2016zud} is formally equivalent to the RT prescription, provides useful tools and visualizations for understanding the structure of entanglement in holographic states.\footnote{The bit threads formalism has further been extended to include multiflows and thread configurations~\cite{Cui:2018dyq}.} In essence, the bit threads formulation of the RT formula replaces minimization of areas of surfaces with maximizations of constrained flows of vector fields through surfaces.
These vector fields are envisioned as providing, at least heuristically, connections between Bell pairs of maximally-entangled qubits in the boundary state. 

The usual conceptual flow of the bit threads program is to pass from a given state to a bit thread geometry describing (aspects of) the entanglement structure of the state.
In this paper---and in the context of a holographic correspondence between boundary states and bulk geometries---we wish to invert this logic: we conjecture that, under certain circumstances discussed below, the existence of a bit thread geometry satisfying particular geometric constraints \textit{implies} the existence of a state with subsystem entanglements satisfying the analogues of these constraints.
Given this conjecture, we can replace the (extremely hard) task of specifying a particular holographic state directly in the boundary CFT with a more tractable geometric construction on the level of a classical bulk geometry. 
Because we expect that geometrical information is encoded redundantly in the boundary theory in the manner of an error-correcting code (see, e.g., Ref.~\cite{Harlow:2016vwg}), we do not expect this construction to uniquely specify a quantum-mechanical state.
Nevertheless, we show in this paper that it provides enough information to compute the information-theoretic quantity dual to the cross-sectional area $E_W$ and hence establish that $E_P=E_W$.

The organization of this paper is as follows. In \Sec{sec:background}, we review some facts about bit threads and the entanglement of purification.
In \Sec{sec:upperbound}, we argue that the entanglement of purification is bounded from above by the area of the entanglement wedge cross section, while in \Sec{sec:lowerbound} we show, under certain assumptions that we clarify, that the entanglement of purification is lower-bounded by the area of the entanglement wedge cross section, establishing the $E_P = E_W$ relation.
Finally, we conclude in \Sec{sec:discussion}, with a bonus related result on the minimal dimension of holographic purifications given in \App{app:mindim}.

\section{Background}\label{sec:background}

We begin by introducing some definitions and briefly reviewing the concepts of entanglement of purification and bit threads.

\subsection{Entanglement of purification}
\label{ssec:EP}

Let $\rho_{AB} \in \mathcal{L}(\Hil_{AB})$ be a state in the bipartite Hilbert space $\Hil_{AB} = \Hil_A \otimes \Hil_B$.
The \emph{entanglement of purification} of $\rho_{AB}$ is
\begin{equation}
E_P(A:B) = \inf_{A^\pr B^\pr} S(AA^\pr),
\end{equation}
where the infimum is taken over all auxiliary Hilbert spaces $A^\pr B^\pr$ and all pure states $\ket{\Psi}_{AA^\pr BB^\pr}$ such that $\Tr_{A^\pr B^\pr} \ketbra{\Psi}{\Psi} = \rho_{AB}$.
The entanglement of purification is an entanglement measure that quantifies the minimum amount of correlation required in any purification of a mixed state, subject to the constraint that it must preserve the factorization structure between $A$ and $B$; this is particularly useful, for example, when only a mixed subset of the full pure state is known.

\begin{figure}[ht]
\centering
\includegraphics[scale=0.5]{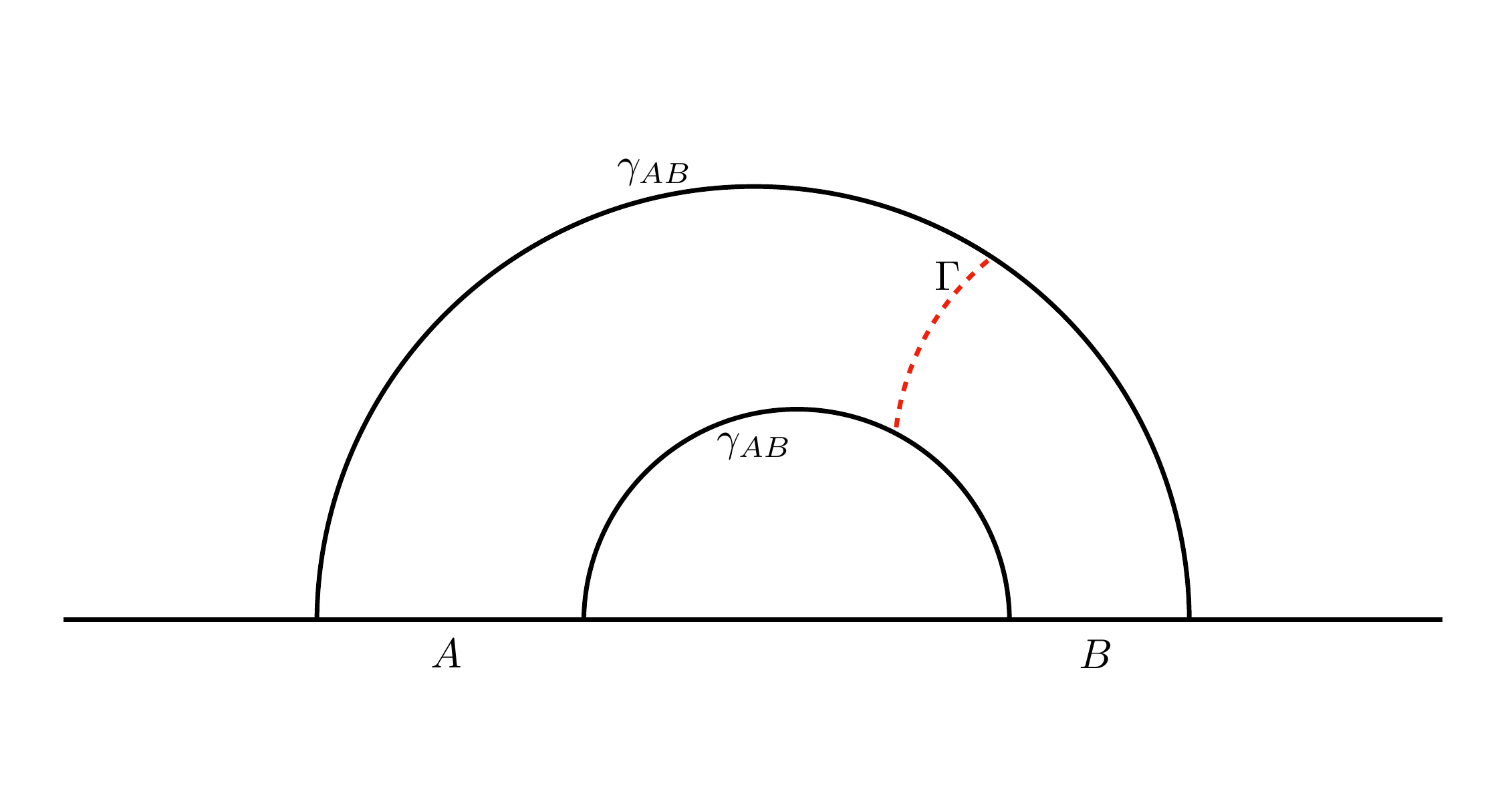}
\caption{Entanglement wedge for $AB$, bounded by the RT surfaces $\gamma_{AB}$. The entanglement wedge cross section $\Gamma$ is illustrated by the red dashed line.}
\label{fig:EW}
\end{figure}

When $\Hil_{AB}$ is a subfactor of the Hilbert space of a holographic CFT and $\rho_{AB}$ is the reduced state on boundary subregions $A$ and $B$ of a state with a well-defined dual geometry, the entanglement of purification $E_P$ has been conjectured to be equal to the area $E_W$ of the entanglement wedge cross section $\Gamma$ \cite{Takayanagi:2017knl}.
The entanglement wedge cross section $\Gamma$ is a surface of minimal area anchored to the boundary of the entanglement wedge $W_{AB}$, such that $\Gamma$ partitions $W_{AB}$ into a region that is entirely adjacent to $A$ and a region that is entirely adjacent to $B$; see \Fig{fig:EW}.

In the case of a finite-dimensional Hilbert space, $\dim \Hil_{AB} = d_{AB} < \infty$, the infimum $E_P(A:B)$ can be achieved with auxiliary Hilbert spaces of dimensions $d_{A'} = d_{AB}$ and $d_{B'} = d_{AB}^2$ \cite{terhal2002entanglement}.
Note that while the Hilbert spaces of AdS/CFT are infinite-dimensional, we will later consider UV regularizations that render them finite-dimensional.
Therefore, we can assume that the infimum is achievable in regulated holographic settings.

The state that achieves the entanglement of purification as $S(AA')$ is of course not unique.
(For example, given a state $\ket{\Psi}$ that achieves $E_P(A:B)$, any other state of the form $U_{A'} \otimes U_{B'}\ket{\Psi}$ also achieves $E_P(A:B)$.)
One should therefore think of entanglement of purification as picking out an equivalence class of ``optimal'' states on Hilbert spaces of possibly different dimensions that are all isometrically related.
In \App{app:mindim}, we show that if one holds a purification of holographic $\rho_{AB}$ in $\Hil_{CFT}$ that achieves $E_P(A:B)$, then one can compress the purifying state on $(AB)^c$ to a Hilbert space with $\log \dim \Hil_{A'B'} = S(\rho_{AB})$.

\subsection{Bit threads}

Bit threads were first introduced as a by-product of a reformulation of the RT formula in terms of flows by Freedman and Headrick in \Ref{Freedman:2016zud}.\footnote{The formulation of the RT prescription in terms of flows can also be understood~\cite{Bakhmatov:2017ihw} from the perspective of calibrations defined on Riemannian manifolds~\cite{Harvey:1982xk}.}
Here, we introduce only the minimum results required for our purposes, leaving the details to \Ref{Freedman:2016zud}.

Let $\mathcal{M}$ be an oriented Riemannian manifold with boundary and let $C > 0$. A \emph{flow} is a vector field $v$ such that $\nabla_\mu v^\mu = 0$ and $|v| \leq C$.
Given a flow, one can of course compute the flow's flux through any sufficiently smooth codimension-one surface $m$:
\begin{equation}
\int_m v = \int_m \sqrt{h} \, n_\mu v^\mu, 
\end{equation}
where $h$ is the induced metric on $m$ and $n$ is the normal to $m$.

The crux of Freedman and Headrick's reformulation of the RT formula is the Max-Flow/Min-Cut (MFMC) Theorem \cite{Federer,Strang1983,MR1088184}, which can be stated as follows.
Let $A$ be a boundary subregion of $\mathcal{M}$. Then
\be 
\max_v \int_A v = C \min_{m \sim A} \mathrm{area}(m),\label{thm:mfmc}
\ee
where $m \sim A$ denotes that $m$ is homologous to $A$.
MFMC states that for any flow on $\mathcal{M}$ that attains the largest possible flux out of $A$, the value of this flux is equal to the largest transverse flow density $C$ multiplied by the area of a minimal surface in $\mathcal{M}$ that is homologous to $A$.
In other words, the max flow necessarily saturates any bottleneck out of $A$ in both magnitude and direction. 

Letting $\mathcal{M}$ be a spatial slice of a static holographic geometry and $C = 1/4G_N\hbar$, it should now not seem surprising that one can rewrite \Eq{eq:RT} as 
\begin{equation}
S(A) = \max_v \int_A v \, .\label{eq:SAflow}
\end{equation}
This was rigorously demonstrated in \Ref{Freedman:2016zud}.

\emph{Bit threads} are the integral curves of a max flow with transverse density $|v|$.
As a result of \Eq{eq:SAflow}, one can think of $S(A)$ as counting the largest number of bit threads leaving $A$ that can pass through the RT surface for $A$.
Due to the non-uniqueness of the max flow, one can also compute further entropic quantities by counting bit threads of appropriately chosen flows.
We will comment further on how this is done in \Sec{sec:upperbound} when we begin our calculations of this type.

\section{Upper Bound on $E_P$}\label{sec:upperbound}

Let $\rho_{AB}$ be a holographic CFT state on the subregions $A$ and $B$ whose geometric dual is the entanglement wedge of $AB$.
In this section, we establish an upper bound on its entanglement of purification, $E_P(A:B)~\leq~E_W(A:B)$, under an assumption regarding the relation between the entanglement structure of geometric purifications of the state and a bit thread flow in the entanglement wedge. 
To do so, it suffices to exhibit a purification for which $S(AA') = E_W(A:B)$. 
We will show that an arbitrary purification that is everywhere holographically dual to an asymptotically-AdS bulk geometry admits a factorization $\ket{\Psi}_{AA'BB'Y}$ for which $S(AA') = E_W(A:B)$.
Such a purification is guaranteed to exist by the fact that geometric subregions of an AdS geometry (in this case the entanglement wedge) can always be extended to a full asymptotically-AdS bulk geometry.\footnote{In the case where $\rho_{AB}$ resulted from tracing out $(AB)^c$ in a holographic state, then one trivial everywhere-geometric purification is the original state.
More generally, on the level of the bulk geometry, a geometric purification corresponds to gluing the entanglement wedge of $AB$ to other subregions in such a way that the entire geometry is asymptotically AdS.
In arbitrary dimensions such a gluing construction may be nontrivial for general geometries, but it has been shown that gluing another copy of the entanglement wedge to itself along RT surfaces is always possible, forming a canonical purification~\cite{Faulkner,EW}; in $\mathrm{AdS_3}$, where the Weyl tensor vanishes identically, even more general gluing constructions should be fairly straightforward.
Note that it may be necessary to introduce additional copies of the CFT if, for example, $AB$ is already a full boundary and $\rho_{AB}$ is a mixed state, such as in the case of a thermal single-sided black hole.
\label{foot:Weyl}}
We now specify how to determine $S(AA')$ for this state and for a particular choice of $A'$, which we fix by using the bit thread formalism to specify various conditional mutual informations. 

\subsection{Bit threads as tags of purifying degrees of freedom}

\begin{figure}[ht]
\centering
\includegraphics[scale=0.5]{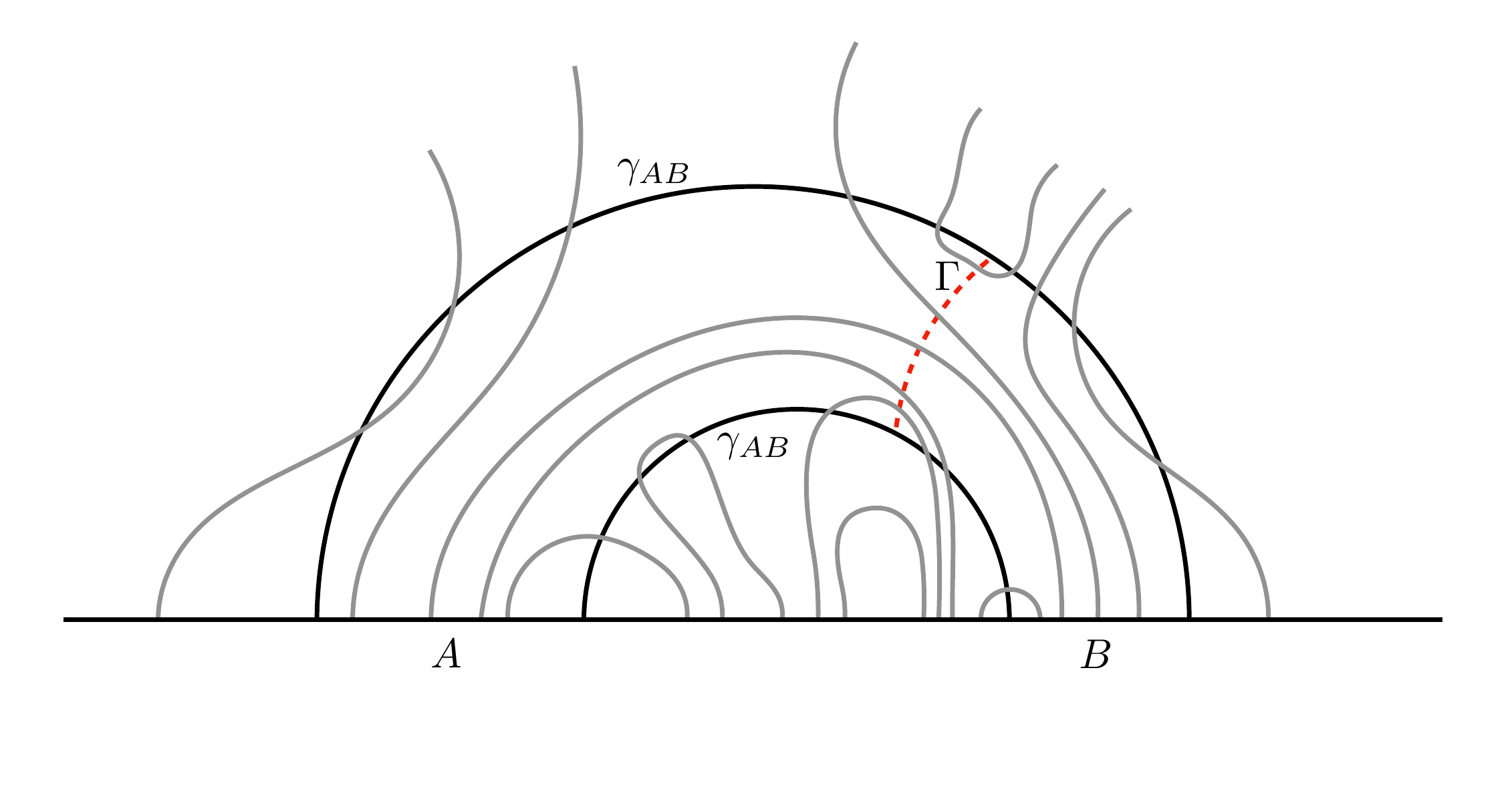}
\caption{Types of bit threads that can cross the RT surface, $\gamma_{AB}$, of $AB$ for an arbitrary flow. The properties that distinguish different types of bit threads are where they begin and end ($A$, $B$, or $(AB)^c$), as well as how many times they intersect $\gamma_{AB}$ and the entanglement wedge cross section $\Gamma$ (red dashed line).}
\label{fig:allthreads_general}
\end{figure}

Consider two boundary subregions $A$ and $B$, as well as the bit threads of some flow on a geometric purification of $\rho_{AB}$.
The ways in which bit threads can intersect the RT surface of $AB$ (which in general consists of multiple disconnected pieces) are sketched in \Fig{fig:allthreads_general}.

\begin{figure}[ht]
\centering
\includegraphics[scale=0.5]{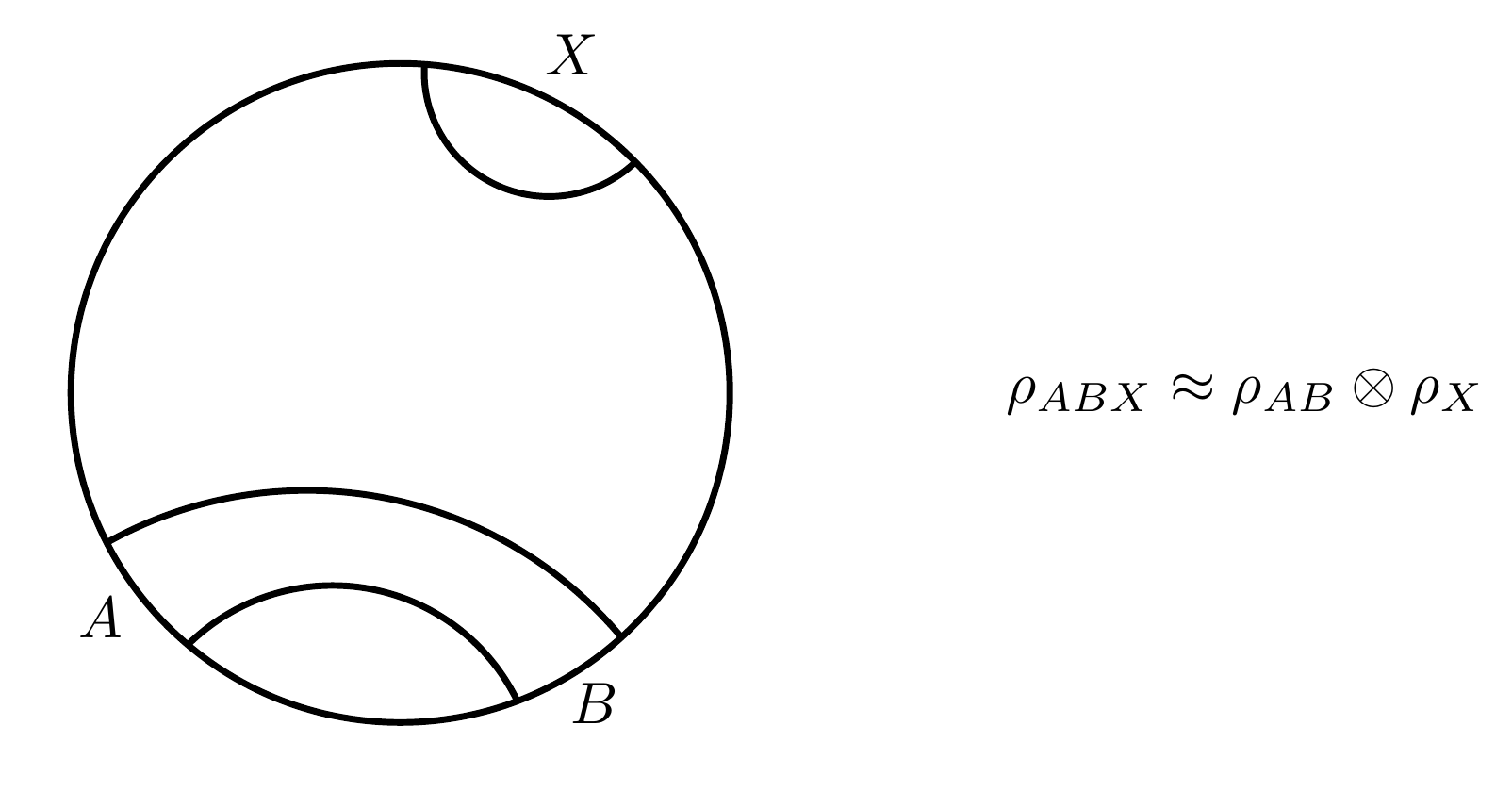}
\caption{Boundary subregions whose reduced density matrix approximately factorizes.}
\label{fig:approx_factor}
\end{figure}

Threads that start in $AB$ and terminate in $(AB)^c$ are to be associated with the subfactor of $\Hil_{(AB)^c}$ that purifies $\rho_{AB}$.
That is, $(AB)^c$ contains many extraneous degrees of freedom that are uncorrelated with $\rho_{AB}$ and hence do not purify our reduced density matrix.
This can be seen, for example, by looking at small boundary subregions $X \subset (AB)^c$ that are far away from $\partial(AB)$, as shown in \Fig{fig:approx_factor}.
Such regions have zero mutual information with $AB$ and the state factorizes\footnote{In field theory, the fact that the vacuum state is entangled at all scales implies that even local operators well outside the entanglement wedge might change the reduced state $\rho_{AB}$. A finite-energy particle excitation at a particular spacetime location in fact has (exponentially small) support on all of space. So strictly speaking, we should throughout replace any operator acting outside the entanglement wedge with a different operator, with cutoff-scale energy, that genuinely does not change the reduced state of the entanglement wedge. Or, perhaps preferably, we could instead not demand that the new reduced state be exactly the same as the old, but only the same up to exponentially small differences. That is, whenever we factorize the Hilbert space of a CFT, we can either replace statements of factorization with appropriate statements involving von Neumann algebras (and/or bulk reconstruction) or edge modes, or we can embrace the existence of the cutoff and work with a latticized theory that lacks these issues. In the remainder of this paper, we take the latter, more pedestrian approach.\label{foot:factor}} in the large-$N$ limit:\footnote{It should be noted that (even beyond the caveats just noted in footnote~\ref{foot:factor}) this factorization is approximate, to leading order in $1/N$, since bit threads are equivalent to the RT formula, which only computes the entropies to leading order in $1/N$. We are helped, however, by the fact that the trace distance between the exact state and the factorized state discussed here will be small, so Fannes' inequality~\cite{Fannes1973} guarantees that the difference in the entanglement entropies of subsystems of the two states will also be correspondingly small.}
\begin{equation}
\rho_{ABX} = \rho_{AB} \otimes \rho_{X}.
\end{equation}

As such, we conclude that it is only a subset of the degrees of freedom, $\Hil_{(AB)'} \subset \Hil_{(AB)^c}$, that purify $\rho_{AB}$; bit threads are to be associated with this smaller set of boundary degrees of freedom. 
Although we have chosen notation here to suggest a division $\Hil_{(AB)'} = \Hil_{A'}\otimes\Hil_{B'}$, so far we have not actually specified such a division, nor in fact required that $\Hil_{(AB)'}$ even factorize.
In particular, $\Hil_{(AB)'}$ need not have any geometric locality, beyond simply that it is supported in $(AB)^c$.
This is a reflection of the freedom in choosing the flow, i.e., the fact that the location where the threads terminate in $(AB)^c$ can be slid around freely anywhere in $(AB)^c$, and so they are not associated with literal, local ``boundary qubits.''
For example, the threads could be evenly spaced out, or we could bunch them up together about a particular location on the boundary, etc.

\subsection{A flux-maximizing thread configuration}

In order to fix the division of $(AB)'$ into $A'$ and $B'$, we now construct a specific configuration of bit threads that simultaneously maximizes the number of threads crossing the RT surface of $AB$, which we denote by $\gamma_{AB}$, and the number crossing the entanglement wedge cross section $\Gamma$.
Note, however, that such a collection of threads cannot be the flow lines of an everywhere-continuous and -divergenceless flow with bounded norm.
According to MFMC \eqref{thm:mfmc}, if a flow $v$ indeed maximizes the flux out of $AB$, then $v$ must be normal to $\gamma_{AB}$ and have $|v| = 1/4G_N\hbar$ everywhere on $\gamma_{AB}$.
But then, the flux through $\Gamma$ cannot be strictly equal to $|\Gamma|/4G_N\hbar$ since, on the codimension-three surface where $\Gamma$ intersects $\gamma_{AB}$, $v$ is perpendicular to the normal of $\Gamma$.

Therefore, we must necessarily relax the definition of bit threads as integral curves of a flow.
We will simply take bit threads to be boundary-anchored one-dimensional objects whose density is at most $1/4G_N\hbar$, defining thread density, as in \Ref{Agon:2018lwq}, as the length of threads within some small neighborhood divided by the volume of that neighborhood.
This generalization was proposed in \Ref{Cui:2018dyq}, in which a collection of such threads was called a ``thread configuration.''
Although the authors of \Ref{Cui:2018dyq} further relaxed the requirement that bit threads be oriented, we will find it helpful to still think of our bit threads as having an orientation.

The use of this more general notion of thread configurations is very mild for our purposes, since ultimately we will only be concerned with computing entropic quantities by counting bit threads, as opposed to exploiting MFMC to identify the bottleneck of a flow.
In other words, we will always assume that the relevant extremal surfaces are known, and we will only need to count threads that cross these surfaces. Moreover, the only place where the associated flow becomes multivalued is on the (codimension-three) surface $\Gamma \cap \gamma_{AB}$, so with bounded flux density, the region of this ambiguity contributes only a measure-zero fraction of the flux through $\Gamma$ or $\gamma_{AB}$.

With this generalization in mind, we can now describe how to construct a thread configuration that saturates the number of threads crossing $\gamma_{AB}$ and $\Gamma$.
Let us suppose that a UV cutoff near the boundary has been imposed, so that the areas of boundary-anchored surfaces are finite and well defined.
First, begin by letting $S(A) = |\gamma_A|/4G_N\hbar$ thread fragments emanate from $A$ and $S(B) = |\gamma_B|/4G_N\hbar$ thread fragments flow into $B$, where $\gamma_A$ and $\gamma_B$ denote the RT surfaces of $A$ and $B$ individually and where the thread fragments run from the boundary to $\gamma_A$ and $\gamma_B$ for now.
These are the largest numbers of threads that we can pipe from $A$ to the bulk and to $B$ from the bulk without violating the density bound anywhere. The reader will note that we have suggestively labeled the numbers of threads using the same symbol as entropy, and we will proceed to work with ``entropies'' and ``mutual informations,'' but, as we discuss in the next subsection, \textit{a priori} these quantities are purely geometric and need not be associated with the entanglement entropy of reduced density matrices constructed from a pure state.

One of our goals is to saturate the number of threads that cross $\gamma_{AB}$, so $S(AB) = |\gamma_{AB}|/4G_N\hbar$ of the fragments will have to cross $\gamma_{AB}$ and leave the entanglement wedge.
Therefore, take $[S(A) + S(B) - S(AB)]/2$ of the thread fragments from $A$ and the same number from $B$ (i.e., the leftover fragments) and join them by having them cross $\Gamma$.
At this stage, there are thus $[S(A)+S(B)-S(AB)]/2 = I(A:B)/2$ bit threads crossing $\Gamma$.

\begin{figure}[!tbph]
\centering
\includegraphics[scale=0.35]{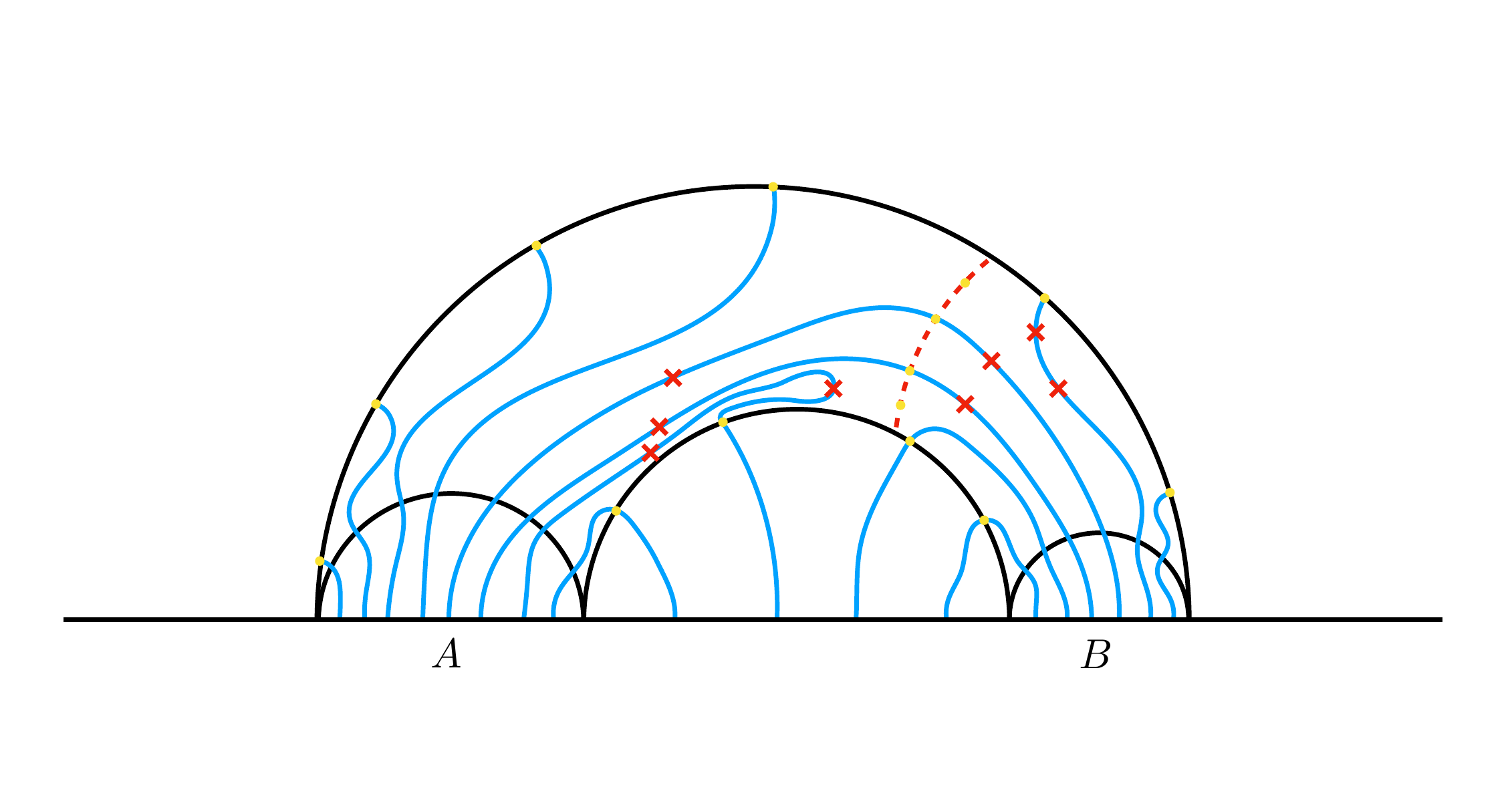}\includegraphics[scale=0.35]{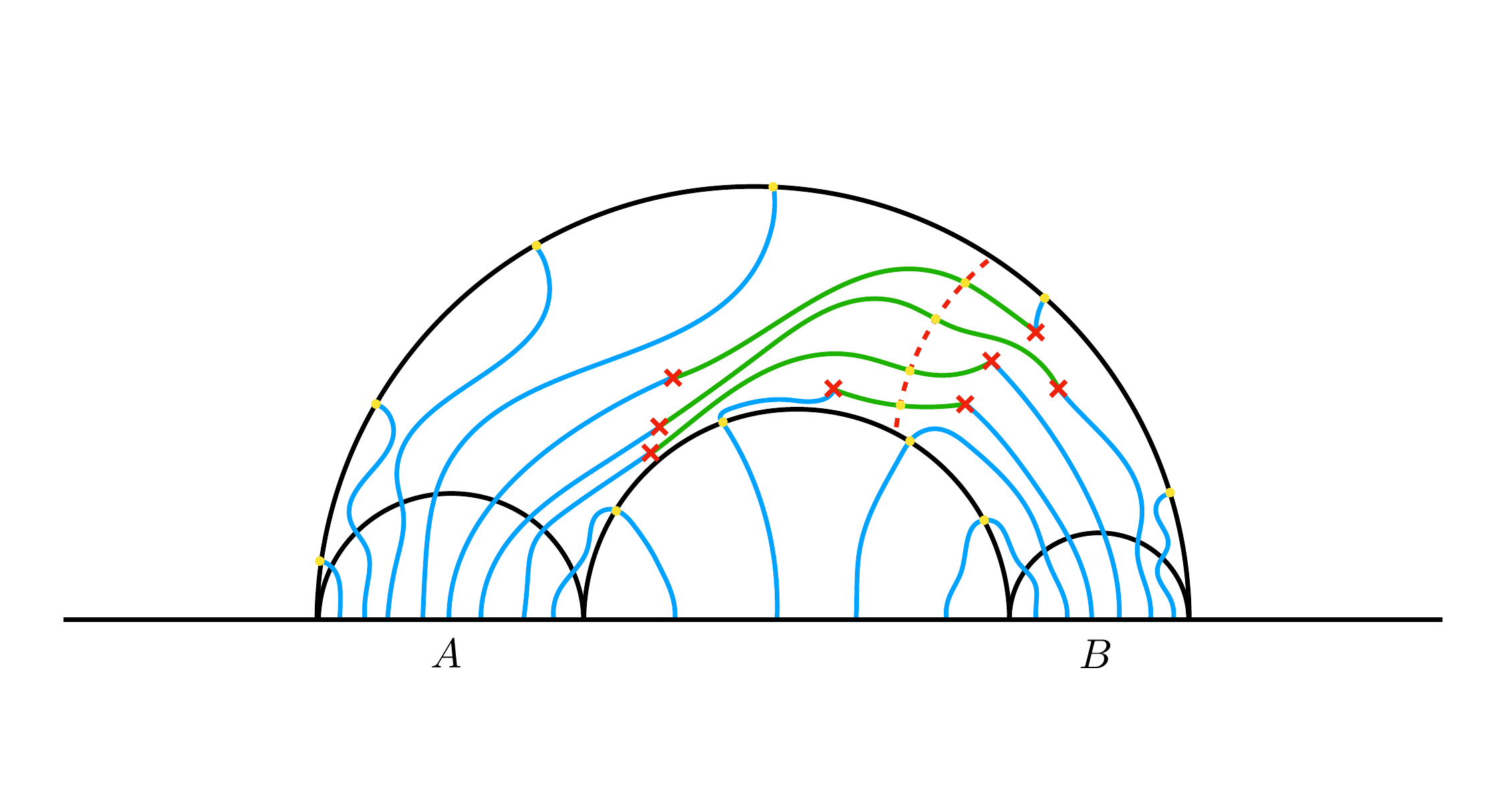}
(a) \hspace{0.45\textwidth} (b)

\vspace{2mm}

\includegraphics[scale=0.35]{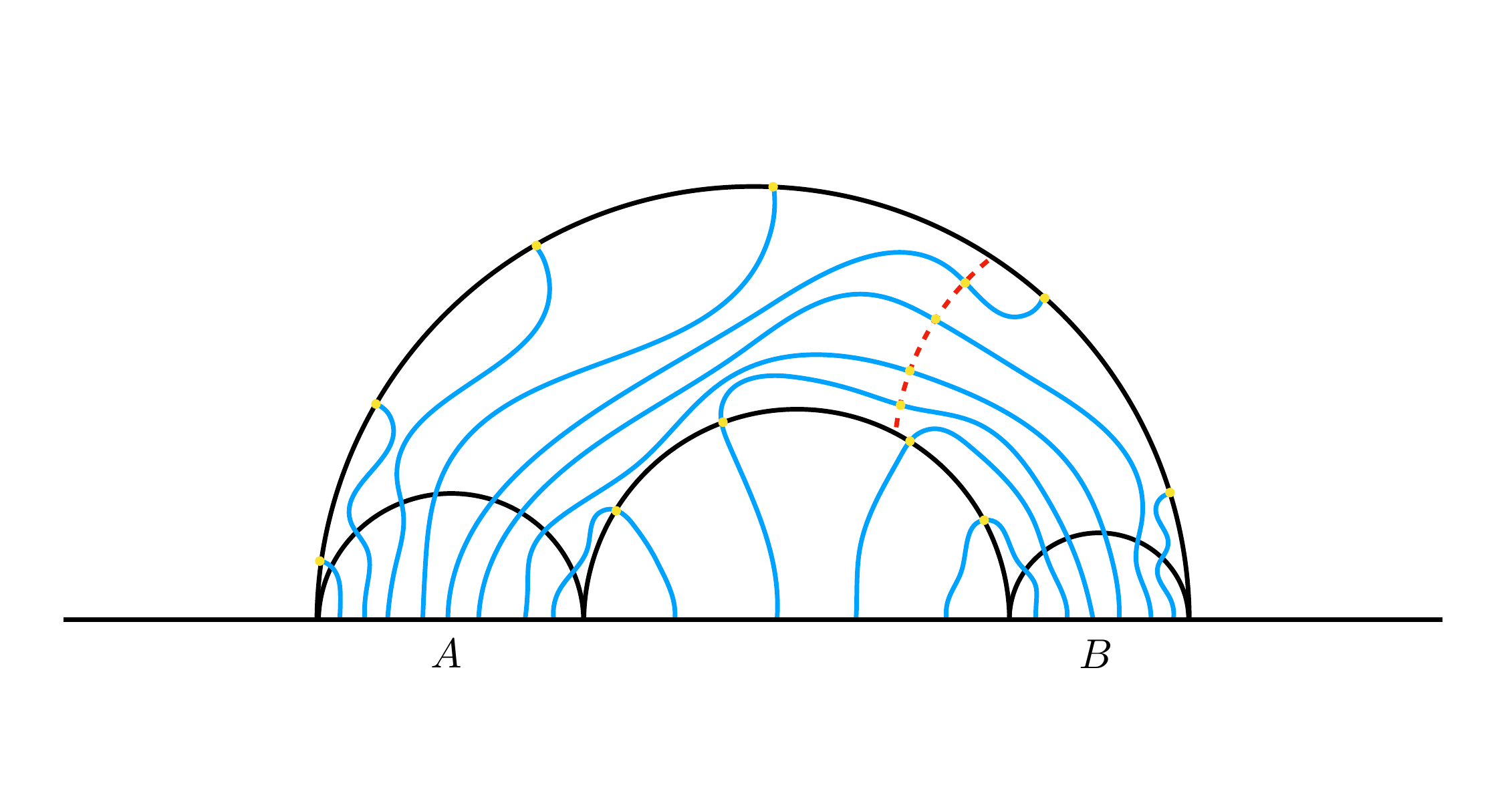}\includegraphics[scale=0.35]{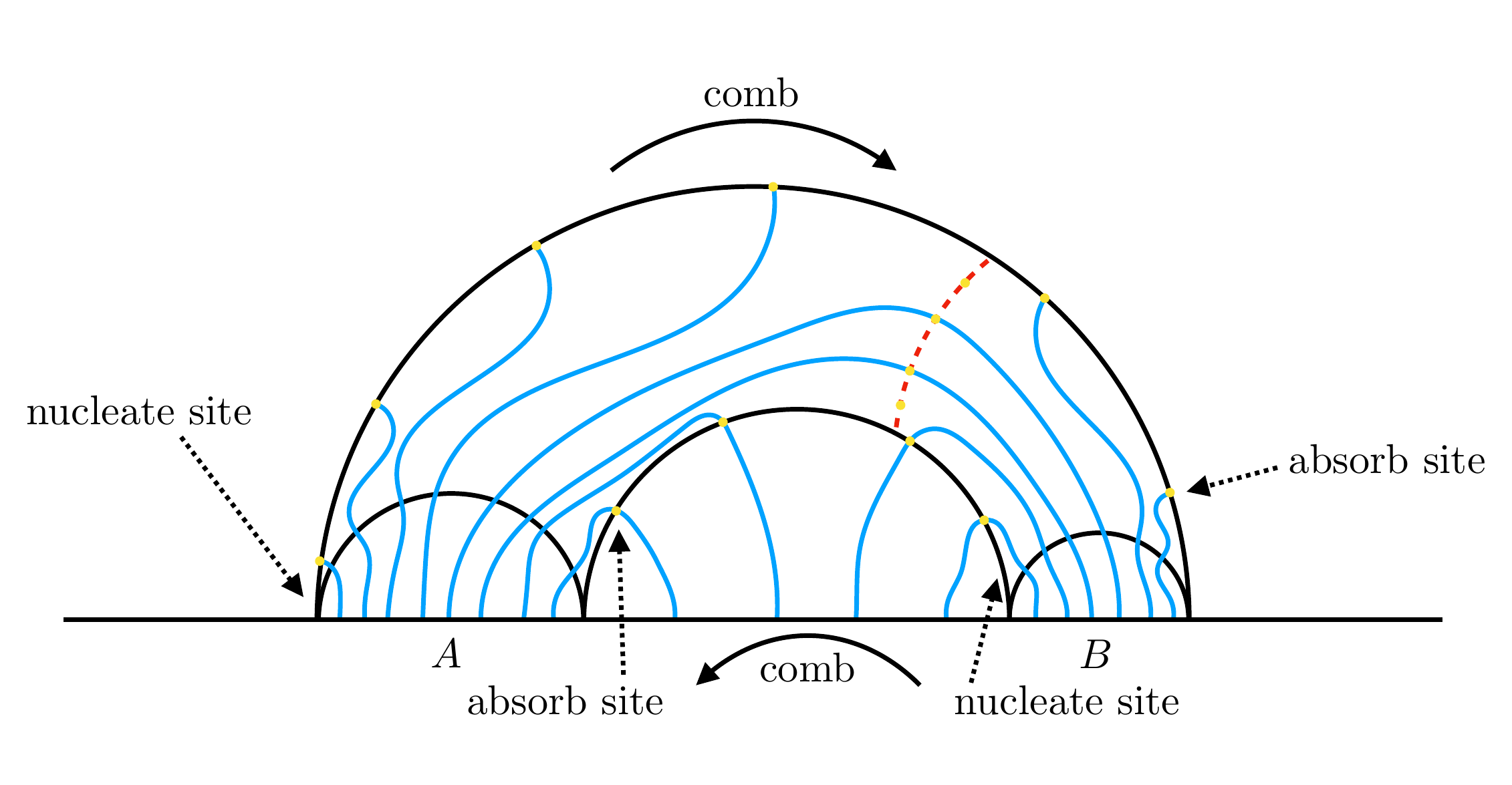}
(c) \hspace{0.45\textwidth} (d)
\caption{Starting with a thread configuration that saturates the number of threads intersecting $\gamma_{AB}$, $\gamma_A$, and $\gamma_B$, as shown in (a) and described in \Ref{Agon:2018lwq}, the goal is to construct a configuration like (c), which also saturates $\Gamma$. (The parts of threads that cross the exterior component of $\gamma_{AB}$ and that lie outside the entanglement wedge have been suppressed in these diagrams.)
Diagram (b) illustrates how this can be done by cutting and gluing threads.
First, with a UV cutoff in place, it is helpful to think of there being a finite number of sites (yellow dots) on both $\gamma_{AB}$ and $\Gamma$ that threads must intersect.
Initially, only $I(A:B)/2$ sites on $\Gamma$ are filled by threads, which form a tube running through $\Gamma$.
Divide the remaining $E_W(A:B) - I(A:B)/2$ sites into a group of $n_{\mrm{hi}}$ sites above the tube and $n_{\mrm{lo}}$ sites below the tube.
(In two spatial dimensions the division is unique, but in higher dimensions there may be freedom in this choice.)
Then, cut the $I(A:B)/2$ threads that cross $\Gamma$, as well as $n_{\mrm{hi}}$ threads that intersect the exterior component of $\gamma_{AB}$ on the side adjacent to $B$ and $n_{\mrm{lo}}$ threads that intersect the interior component of $\gamma_{AB}$ on the side adjacent to $A$.
The locations of the cuts are indicated by red crosses in (a).
Finally, glue adjacent cut threads together as shown in (b) to arrive at the maximizing configuration (c).
This cutting-and-gluing procedure is equivalent to ``combing'' the original thread configuration, as depicted in (d).
Combing means dragging $n_{\mrm{hi}}$ threads that intersect sites on the $A$-adjacent side of the exterior component of $\gamma_{AB}$ over to sites on the $B$-adjacent side, and vice-versa for $n_{\mrm{lo}}$ sites on the interior component of $\gamma_{AB}$.
The only additional subtlety is that, with a UV cutoff in place, we must think of $n_{\mrm{hi}}$ threads being nucleated from the UV where the exterior component of $\gamma_{AB}$ meets $A$ and $n_{\mrm{hi}}$ sites being absorbed by the UV where the exterior component of $\gamma_{AB}$ meets $B$.
The same is true for $n_{\mrm{lo}}$ threads hitting the interior component, with $A$ and $B$ flipped.
Although the diagrams as we have drawn them here are directly reflective of AdS\textsubscript{3}/CFT\textsubscript{2}, we do not believe that there are any barriers to combing and cutting-and-gluing in arbitrary dimensions.
}
\label{fig:cutandpaste}
\end{figure}

Next, since $E_W(A:B) \geq I(A:B)/2$ (as shown in \Refs{Freedman:2016zud,Takayanagi:2017knl}), we may still need to pipe some of the thread fragments through $\Gamma$.
This is just a matter of ensuring that $E_W(A:B) - I(A:B)/2$ of the remaining $S(AB)$ fragments cross $\Gamma$ before going on to intersect $\gamma_{AB}$, after which the threads are sent to the boundary where they terminate in $(AB)^c$.
 In particular, we can see that there are enough remaining threads to saturate $\Gamma$ as follows:
\begin{equation}
\begin{aligned}
S(AB) + I(A:B)/2 &= \tfrac{1}{2}\left[S(A) + S(B) + S(AB)  \right]\\
&\geq \max\{S(A),S(B)\} \\
&\geq E_W(A:B).
\end{aligned}
\end{equation}
In going to the second line we used the Araki-Lieb inequality, and to go to the third line we used the inequality $E_W(A:B) \leq \min\{S(A),S(B)\}$ \cite{Takayanagi:2017knl}.
The result is a thread configuration with $S(AB)$ threads crossing $\gamma_{AB}$ and $E_W(A:B)$ threads crossing $\Gamma$, thus saturating both surfaces as desired (see \Fig{fig:cutandpaste}c).

Alternatively, we can obtain our $\gamma_{AB}$- and $\Gamma$-saturating thread configuration directly from the thread configuration constructed in Sec.~5.2.2 of \Ref{Agon:2018lwq}, via either cutting-and-gluing or combing.
There, Ag{\'o}n et al. construct a configuration of threads that saturates the number of threads crossing $\gamma_{AB}$, but not $\Gamma$ (see their Fig.~11).
From their configuration, one can arrive at a new configuration that saturates both quantities by cutting threads on either side of $\Gamma$ and rejoining them with neighboring threads inside the entanglement wedge to pipe $E_W(A:B)$ threads through $\Gamma$; this is illustrated in \Fig{fig:cutandpaste}a-b.
This is equivalent to ``combing'' the thread configuration by dragging the threads that intersect $\gamma_{AB}$ in opposite directions along the interior and exterior disconnected components of $\gamma_{AB}$, as illustrated in \Fig{fig:cutandpaste}d. 
We refer the reader to the caption of \Fig{fig:cutandpaste} for more details.

\begin{figure}[ht]
\centering
\includegraphics[scale=0.6]{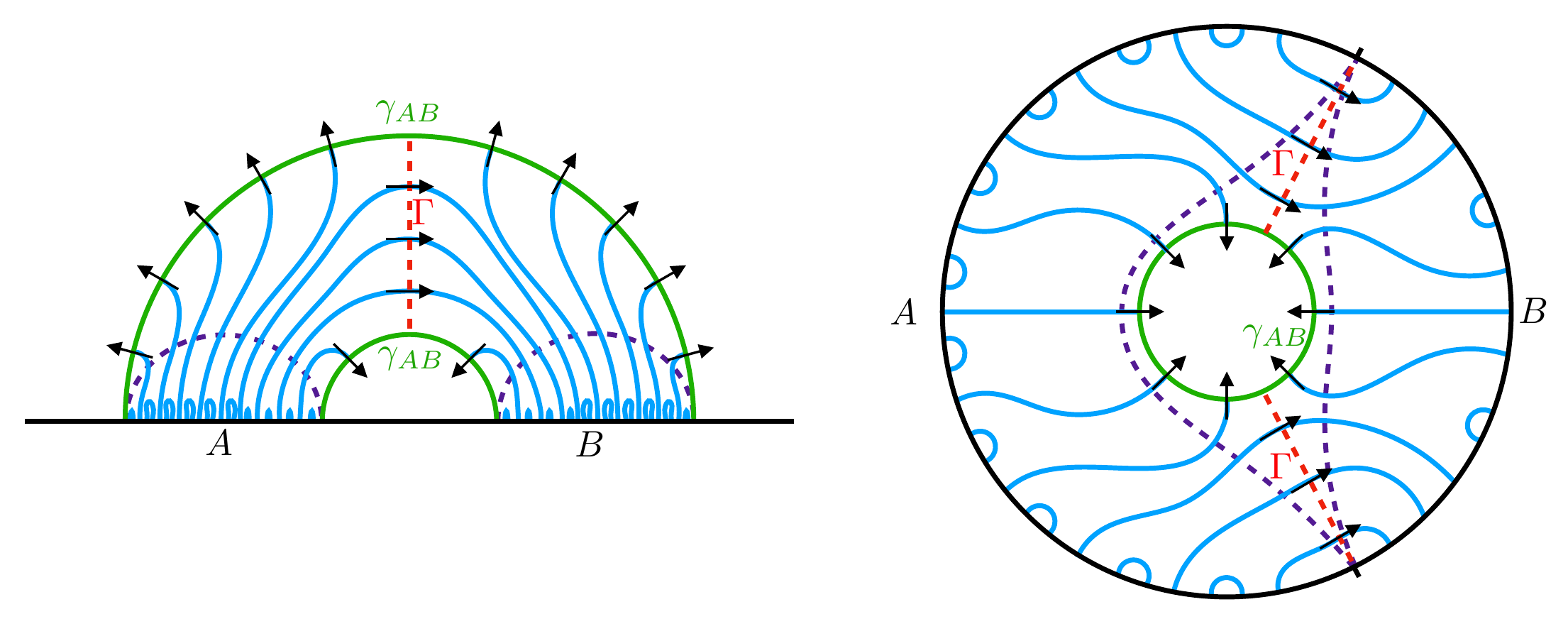}
\caption{Representative examples of bit thread configurations simultaneously maximizing the flow through the entanglement wedge surface $\Gamma$ (red dashed line) and RT surface $\gamma_{AB}$ (green line). The vector fields specifying the flow are indicated on $\gamma_{AB}$ and $\Gamma$ by arrows, and the individual RT surfaces for $A$ and $B$ are depicted with purple dashed lines. Left: $A$ and $B$ are two surfaces defining a proper subregion of a single boundary, on which the CFT state is pure. Right: $A$ and $B$ partition an entire boundary, on which is defined a mixed CFT state, resulting in a horizon in the bulk.}
\label{fig:examples}
\end{figure}

Ag\'on et al.'s construction applies specifically to the case where $AB$ is a proper subset of a single boundary CFT and $W_{AB}$ is simply connected.
Nevertheless, a thread configuration that maximizes the number of threads intersecting $\gamma_{AB}$ and $\Gamma$ can always be constructed according to the first prescription that we gave above.
This is because the inequalities $S(AB) \leq S(A) + S(B)$, $I(A:B)/2 \leq E_W(A:B) \leq \min\{S(A),S(B)\}$, and $S(AB) \geq |S(A)-S(B)|$ guarantee that the right number of thread fragments can always be connected across $\gamma_{AB}$ or $\Gamma$ and attached to $A$ and $B$ as we described to construct the maximizing configuration.
A qualitatively different example is the case where $W_{AB}$ is disconnected, i.e., $\gamma_{AB}$ is just the union of $\gamma_A$ and $\gamma_B$.
In this case $E_W(A:B) = 0$ and so there is no surface $\Gamma$ to saturate.
Another example is when $AB$ is an entire CFT boundary but $\rho_{AB}$ is a mixed state, such as the case of a single-sided mixed state black hole shown in \Fig{fig:examples}.
In this latter example, threads that leave $W_{AB}$ terminate on the black hole horizon.\footnote{Alternatively, we can think of the threads as crossing through a wormhole and terminating in another asymptotic boundary in which the black hole is purified.}

Regulating lengths of surfaces near the boundary using a cutoff $\epsilon$ at finite coordinate distance into the bulk, the length of a generic boundary region in Planck units scales as $\epsilon^{-1}$, while boundary- and bulk-anchored surfaces scale as $\log \epsilon$ and $\epsilon^0$, respectively. Thus, there will always be a great excess of available bit threads anchored to $A$ and $B$ to saturate the flow through the RT and entanglement wedge surfaces. Viewing the bit thread configurations as integral curves specified by a choice of vector field in the bulk, we must guarantee that the fields are chosen orthogonal to both the entanglement wedge cross section $\Gamma$ and the RT surface $\gamma_{AB}$. As mentioned above, this is always possible except at $\Gamma\cap\gamma_{AB}$, which introduces errors only of measure-zero in the net flux. See \Fig{fig:examples} for representative examples.

\begin{figure}[ht]
\centering
\includegraphics[scale=0.5]{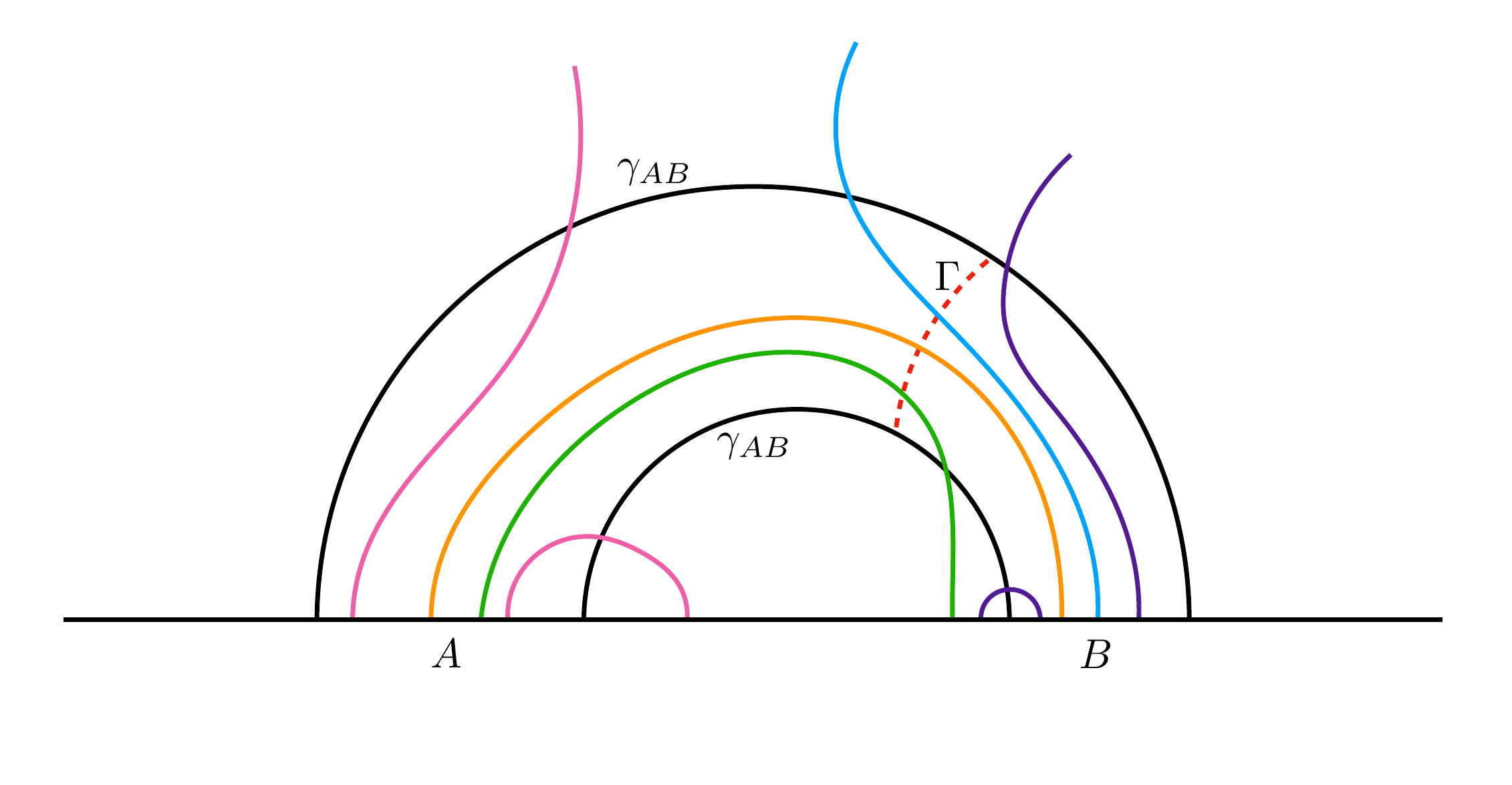}
\caption{The types of bit threads that can cross the RT surface of $AB$ for the configuration that maximizes the flow out of $AB$ as well as the flux through the entanglement wedge cross section $\Gamma$ (red dashed line). We have highlighted threads that have one end in $A$ or $B$ and cross both $\Gamma$ and the RT surface in green or blue, respectively.}
\label{fig:allthreads}
\end{figure}

The possible types of bit threads in a $\gamma_{AB}$- and $\Gamma$-saturating configuration are shown in \Fig{fig:allthreads}.
Here, we have only indicated threads that intersect $W_{AB}$.
Although for a general thread configuration like that of \Fig{fig:allthreads_general} we could have threads that cross the RT surface multiple times, we can without loss of generality take such threads to be absent in a configuration (like that of \Fig{fig:allthreads}) that obeys the maximization condition.
This can be understood as follows.
In the maximum-flux configuration, $|\gamma_{AB}|/4G_N\hbar$ threads cross $\gamma_{AB}$ and are anchored to $AB$.
Bit threads that pass through $\gamma_{AB}$ an even number of times contribute zero net flow out of $\gamma_{AB}$, while those that pass through an odd number of times contribute the same net flow as those that pass through once.
Hence, we can take the threads that pass through $\gamma_{AB}$ to do so exactly once, without decreasing the maximum net flow out of $AB$.
By the same token, since we are considering bit threads that simultaneously  saturate the entanglement wedge cross section $\Gamma$, we can take the bit threads to cross $\Gamma$ at most once, and there are $E_W(A:B)$ bit threads that do so.

Moreover, threads that start on $A$ and end on $A$ (or start on $B$ and end on $B$) are not especially meaningful for our analysis.
According to the maximization condition, there must be $|\gamma_{AB}|/4G_N\hbar$ threads passing through $\gamma_{AB}$ and $E_W(A:B)$ threads passing through $\Gamma$.
However, since $|AB|$ can be large compared to $E_W$ and $|\gamma_{AB}|$ (but finite, once a UV cutoff is in place), it is possible to include some additional threads that start and terminate on $A$ or $B$, while still retaining our max-flow conditions.
However, such threads would necessarily cross either $\gamma_{AB}$ or $\Gamma$ an even number of times (including zero), and hence would contribute nothing to the thread configuration that we are maximizing.
That is, such threads will not be relevant for any of the mutual informations we consider in our later calculations, so we suppress such extraneous threads from all subsequent figures and discussion.

Thus, we can take all bit threads that have some segment within the entanglement wedge to have an end in $A$, an end in $B$, or both.
We will not depict possible threads that do not enter the entanglement wedge at all; according to the above discussion these should be associated with the subfactor $\Hil_{(AB)^c} / \Hil_{(AB)'}$.

\subsection{Factorization and derivation of the upper bound}

As reviewed above, $E_W(A:B)$ can be computed simply by counting the number of threads, in a $\Gamma$-saturating configuration, that cross the entanglement wedge cross section.
Such threads can be categorized by whether they remain within the entanglement wedge connecting $A$ and $B$, or instead start in $A$ and leave through the RT surface $\gamma_{AB}$, or finally terminate in $B$ having come through $\gamma_{AB}$.
If, as suggested above, we identify threads that leave $\gamma_{AB}$ as tags of the degrees of freedom in $(AB)'$, which purify $\rho_{AB}$, then the counting of threads that leave $\gamma_{AB}$ computes various conditional mutual informations among $A$, $B$, and $(AB)'$.

Conditional mutual information (CMI) is a three-party entropic measure, defined as
\begin{equation}
I(A:B|C) = S(AC) + S(BC) - S(C) - S(ABC) \, .
\end{equation}
Similarly to how the number of orange-type threads connecting $A$ and $B$ in \Fig{fig:allthreads} gives $I(A:B)/2$, more generally, counting threads computes CMIs \cite{Freedman:2016zud}.
For example,
\begin{align*}
\tfrac{1}{2} I(A:(AB)'|B) &= \tfrac{1}{2} \left[S(AB) + S((AB)'B) - S(B) - S(AB(AB)')\right] \\[2mm]
&= \tfrac{1}{2} \left[S(AB) + S(A) - S(B)\right] \\[2mm]
&= S(A) - \tfrac{1}{2} I(A:B) \\[2mm]
&= (\text{\# magenta, orange, green threads}) - (\text{\# orange threads}) \\[2mm]
&= (\text{\# magenta, green threads}) \, .
\end{align*}
In the manipulations above, we used the fact that the geometric purification $\ket{\Psi}_{AB(AB)'}$ is pure\footnote{Earlier, we defined the geometric purification $\ket{\Psi}_{AB(AB)^c}$ on one or more full boundaries with $\Hil_{(AB)^c} \cong \Hil_{(AB)'}\otimes \Hil_Y$, but because $Y$ has no entanglement with $AB$, $\rho_{AB(AB)'}$ is pure and hence can be associated with a state vector.} to go from the first line to the second line.

Our crucial assumption will be to now postulate a factorization of $(AB)'$ into $A'$ and $B'$ according to whether threads that intersect $\gamma_{AB}$ also intersect $\Gamma$ and whether they are anchored to $A$ or $B$.
In other words, we suppose that $(AB)'$ factorizes into $A'$ and $B'$ such that, referring to \Fig{fig:allthreads}, the CMIs among $A$, $B$, $A'$, and $B'$ are given by the following thread counts:
\begin{itemize}
\item The total number of threads (crossing $\Gamma$) that pass from $A$ to $B$ counts $I(A:B)/2$.
\item The number of green threads, which start in $A$, cross $\Gamma$, and leave the RT surface, counts $I(A:B'|B)/2$.
\item The number of blue threads, which start in $B$, cross $\Gamma$, and then leave the RT surface, counts $I(B:A'|A)/2$.
\item The number of magenta threads, which start in $A$ and then leave the RT surface without passing through $\Gamma$, counts $I(A:A'|B)/2$.
\item The number of purple threads, which start in $B$ and leave the RT surface without passing though $\Gamma$, counts $I(B:B'|A)/2$.
\end{itemize}
We take these identifications to \emph{define} $A'$ and $B'$.
A priori it is not clear that, for any choice of state $\rho_{AB}$, there always exists a pure state $\ket{\Psi}_{ABA'B'}$ whose CMIs are in fact equal to the numbers of colored threads as described above in a given bit thread configuration. 
In the broader case, with $I(A:B'|B)/2=c_1$ and $I(B:A'|A)/2=c_2$ for arbitrary $c_1$, $c_2$, such pure states probably do not exist in general.
In this section we \emph{assume} that when $\rho_{AB}$ is dual to an entanglement wedge and the purification is geometric, such a factorization of $(AB)'$ into $A'$ and $B'$ always exists.

Another way of phrasing our assumption is as follows.
Let $\Hil$ be the Hilbert space of the full CFT and $\ket{\Psi} \in \Hil$ be geometric.
For any boundary subregion $R$, we can factorize $\Hil = \Hil_R \otimes \Hil_{R^c}$.
There is always a state-dependent factorization $\Hil_{R^c} \cong \Hil_{R'} \otimes \Hil_Y$ such that $\ket{\Psi} = \ket{\psi}_{RR'} \otimes \ket{\zeta}_Y$.
In this example, $\ket{\psi}_{RR'}$ corresponds to $\ket{\Psi}_{AB(AB)'}$ above.
The state dependent factorization in question is precisely that of a partial entanglement distillation, which yields a factorization of exactly this form.
The nontrivial assumption is that $(AB)'$ further factorizes to match the CMIs as counted by bit threads.

Specifically, we can consider ${\cal H}_{A'B'}$ as a subspace of ${\cal H}_{(AB)^c}$ because the bit threads pick out a subset of $(AB)^c$ that is entangled with $AB$, so there should exist a subspace of $(AB)^c$ that is unentangled with $AB$. This subspace will not be geometric, as any geometric subset of $(AB)^c$ will have some entanglement with $AB$, but there is no requirement in our construction that $(AB)'$ be a geometric boundary subregion, only that its support be contained within $(AB)^c$. Therefore, there exists a unitary transformation acting on $(AB)^c$ that takes the state $\rho_{(AB)^c}$ to $\rho_{A'B'}\otimes \ket{\Phi}\bra{\Phi}$, where $\ket{\Phi}$ is a state in a subspace of $\Hil_{(AB)^c}$ representing the degrees of freedom that are unentangled with $AB$.

The best way to visualize this is as a partial entanglement distillation or an entanglement distillation with a partial inversion. One can first distill the entanglement between $AB$ and $(AB)^c$; this will yield a product state of Bell pairs between a subset of $AB$ and $A'B'$, and two unentangled pure states. Now, one can simply invert the half of the entanglement distillation on the $AB$ portion to recover $\rho_{AB}$. As in \App{app:mindim}, it is necessary for the entanglement cost of $\rho_{AB}$ to equal the distillable entanglement of $\rho_{AB}$, as is guaranteed in \Ref{Bao:2018pvs}. This form of partial entanglement distillation is made possible holographically primarily because of the holographic sandwiching result of \Ref{HSW}. This (clearly state-dependent) procedure will then yield the state described in the previous paragraph.

The existence of a bit thread configuration might be interpreted as providing evidence that the state-dependent sub-factorization $(AB)' = A'B'$ can be achieved in the holographic case, but it would be nice to have a more concrete construction of such systems.
We note that the existence of a surface-state correspondence \cite{miyaji2015surface} would immediately imply that such systems exist: in this case we could directly identify local purifying degrees of freedom living on the RT surface.
We discuss this point further in \Sec{sec:discussion} below.

With these caveats, since $E_W$ is given by the number of threads that cross the entanglement wedge cross section, our assumption then implies that
\begin{equation}
E_W(A:B) = \tfrac{1}{2} \left[ I(A:B) + I(A:B'|B) + I(B:A'|A) \right]. \label{eq:threadcount}
\end{equation}
It is also true for general pure states $\ket{\Psi}_{ABA'B'}$ that
\begin{equation}
I(A:B) + I(A:B'|B) + I(B:A'|A) = 2 S(AA') - I(A':B').
\end{equation}
However, just as $I(A:B)/2$ is given by the number of threads connecting $A$ to $B$ in \Fig{fig:allthreads}, $I(A':B')/2$ must be identified with threads passing from one portion of $(AB)^c$ to another. 
But we argued above that these threads do not contribute to the purification of $\rho_{AB}$; they live in the $Y$ factor of the purifying geometry, i.e., in $\Hil_{(AB)^c} / \Hil_{A'B'}$.
Hence, without loss of generality, we may take any bit thread configuration that saturates the RT surface and remove any bit threads that do not enter $W_{AB}$.

Therefore, it is true here that
\begin{equation}
I(A:B) + I(A:B'|B) + I(B:A'|A) = 2 S(AA').
\end{equation}
Inserting \Eq{eq:threadcount}, we immediately have that
\begin{equation}
E_W(A:B) = S(AA')\label{eq:EWSAA'}
\end{equation}
for the choice of $A'$ defined by the bit thread configuration we have specified.
By the definition in \Sec{sec:background}, $E_P(A:B) \leq S(AA')$ for any particular purification, so we have 
\be
E_P(A:B) \leq E_W(A:B). \label{eq:upperbound}
\ee

\section{Lower Bound on $E_P$}\label{sec:lowerbound}

Let us take our entanglement wedge for $AB$ and write a purification as $\ket{\psi}_{AA'BB'Y}$, where we have padded with enough unentangled ancillae that the dimension of $\ket{\psi}$ has the dimension of a full boundary CFT with some cutoff.
In particular, let us choose $\ket{\psi}$ to be a state that achieves the infimum of $S(AA')$. (Once a cutoff is in place, which makes the boundary theory finite-dimensional (cf. \Sec{ssec:EP}), the infimum may always be achieved.)
Now, the set of CFT states $\ket{\chi_i}$ on the entire boundary that are dual to a classical holographic bulk form an overcomplete basis for all CFT states \cite{Botta-Cantcheff:2017gys}.
By a slight generalization, we should be able to express $\ket{\psi}$ as some superposition
\begin{equation}
\ket{\psi} = \sum_{i=1}^M \alpha_i \ket{\phi_i},\label{eq:decomp}
\end{equation}
where the $\ket{\phi_i}$ are a subset of the $\ket{\chi_i}$ for which the reduced density matrix on $AB$ is fixed, $\Tr_{(AB)^c}\ketbra{\phi_i}{\phi_i} = \rho_{AB}$ for all $i$.
That is, we have written the purification as a superposition over entire classical bulk geometries dual to $\ket{\phi_i}$, such that each has the same geometry in the entanglement wedge.

We can argue for the existence of the decomposition~\eqref{eq:decomp} as follows. Starting from some particular geometric purification of $\rho_{AB}$, we are free to add pairs of black holes connected by wormholes\footnote{Altering an asymptotically-AdS geometry by producing an entangled pair of black holes changes the topology of the spacetime. This is certainly an allowed process in quantum gravity~\cite{Garfinkle:1990eq,Maldacena:2013xja,Bao:2015nqa,Giddings:2009gj}; for example, we can consider inserting initial excitations that collide to produce a pair of entangled black holes, i.e., the time-reversal of the process by which a pair of entangled black holes evaporate. Hence asymptotically-AdS spacetimes in which this topology-changing process takes place should still have a good dual description in the large-$N$ CFT; they can be produced simply by specifying appropriate initial conditions, with no need to deform the theory. However, the topology change is a fundamentally nonperturbative process from the point of view of a classical bulk description as a curved-space QFT, and thus we should not expect this new spacetime to be in the same code subspace \cite{Harlow:2016vwg} as one in which the process never occurs.} to the bulk geometry outside of $W_{AB}$. As this is a local bulk modification on only the purifying subsystem (up to the caveats discussed in footnote~\ref{foot:factor}) this action modifies the boundary state without changing the fact that it is a purification of $\rho_{AB}$. (In the case where $AB$ comprises an entire CFT boundary, with thermal density matrix dual to a black hole in the bulk, this construction simply amounts to gluing on different multiboundary wormholes behind the horizon.) Let us call the dual boundary states formed by this procedure $\ket{\xi_j} \subset \{ \ket{\phi_j}\}$, i.e., these geometries are a subset of those given by the set of all geometric purifications of $\rho_{AB}$.\footnote{As noted in footnote~\ref{foot:Weyl}, this black hole gluing construction is straightforward in asymptotically-${\rm AdS}_3$ geometries, due to the vanishing of the Weyl tensor. More generally, the construction of Engelhardt and Wall \cite{EW} (see also \Ref{Faulkner}) allows us to glue black hole handles onto geometries in higher dimensions; though such states may not be dynamically stable~\cite{Witten:1999xp}, this will not matter for our purposes here, since we merely require the existence of such a geometry as a locally time-symmetric solution of Einstein's equations on a Cauchy slice.}
 Importantly, although general purifications need not all live in the same Hilbert space, the particular dual states $\ket{\xi_j}$ all live in a single Hilbert space, that of a large-$N$ holographic CFT, which also contains the geometric purification defined in the previous subsection. (In the case where $\rho_{AB}$ describes a thermal state, the Hilbert space is instead that of two CFTs, and includes the appropriate thermofield double states.)

In particular, we are free to choose the masses and entanglement structure of such black holes, so long as we do not change the purity of the overall state. These black holes should have support over any complete basis of $\Hil_{(AB)^c}$, by the same reasoning as in the case of thermofield double states of differing temperatures~\cite{Cardy:1986ie,Maldacena:2001kr}, as we can include all black holes from those of minimal size to ones that fill up a large portion of the purifying bulk region. 
Concretely, considering a basis $\ket{\omega_k}$ for $\Hil_{(AB)^c}$ and writing $\Tr_{AB} \ketbra{\xi_j}{\xi_j} = \sigma_j$, for all $j$ we should be able to write
\begin{equation}
\sigma_j =\sum_{k} c_{jk}\ketbra{\omega_k}{\omega_k} + \sum_{k\neq k'} c_{jkk'}\ketbra{\omega_k}{\omega_{k'}},
\end{equation}
where the second sum containing the off-diagonal terms is exponentially suppressed relative to the first sum over the diagonal terms,\footnote{We expect this block-diagonal assumption to be true at leading order in $N$ by the eigenstate thermalization hypothesis~\cite{Srednicki}.} and $c_{jk}$ is (to good approximation) invertible.\footnote{Note in particular that we are not claiming that the $\ket{\xi_j}$ form a complete basis for the entire CFT Hilbert space; in particular, we will not be able to reach states that have reduced density matrices on $AB$ orthogonal to $\rho_{AB}$.}
We expect distinct $\sigma_j$, which correspond to distinct classical metrics outside of $W_{AB}$, to be orthogonal up to exponential corrections, i.e., $\Tr \sigma_j \sigma_{j'} \simeq \delta_{jj'}$ (by the same reasoning that implies that distinct thermofield double states have exponentially small overlap~\cite{Almheiri:2016blp}).
Hence, a superposition of the $\ket{\xi_j}$ will retain the same reduced density matrix $\rho_{AB}$ on $AB$.
The linear independence of the $c_{jk}$ weighting allows the $\ketbra{\omega_k}{\omega_k}$ themselves to be isolated by performing Gauss-Jordan elimination over the black hole geometries: $\ketbra{\omega_k}{\omega_k} \simeq \sum_{j} a_{jk}\sigma_j$, where $a = c^{-1}$. Hence, the number of linearly independent black hole states among the $\ket{\xi_j}$ is at least $\dim \Hil_{(AB)^c}$, so for some $\alpha_j$ we have
\begin{equation}
\ket{\psi}=\sum_{j} \alpha_j \ket{\xi_j},
\end{equation}
and therefore the $\ket{\xi_j}$ are a valid choice of $\ket{\phi_i}$ in the sum in \Eq{eq:decomp}.

It is worth emphasizing at this point that this argument suggests a subregion version of the fact that geometric states form an overcomplete basis of the entire Hilbert space. This idea is an intuitive generalization, but we have motivated it above as we are unaware of this statement's appearance in extant literature.

Now we wish to compute $S(AA')$ in $|\psi\rangle$ and determine its minimum value.
Let us write $S_i = {\Tr}_{(AA')^c} \ketbra{\phi_i}{\phi_i}$, so $S_i$ is the entanglement entropy of $AA'$ in the geometry $i$.
In each individual geometry, we can write down a concrete bit thread configuration, where each bit thread is boundary-anchored, so we can compute $S_i$ by counting bit threads.
As shown in \Ref{Almheiri:2016blp}, we may write $S(AA')$ as a sum $\sum_i |\alpha_i|^2 S_i$ under certain assumptions, which we will discuss below.
Thus, we can find a lower bound on $S(AA')$ by simply lower-bounding $S_i$.
We can lower-bound $S_i$ using a bit thread configuration in an everywhere-classical geometry.
Putting this all together, we have:
\be 
\begin{aligned}
E_P(A:B) &= S(\Tr_{BB'Y} \ketbra{\psi}{\psi}) \\
&= \sum_{i=1}^M |\alpha_i|^2 S_i + S_{\rm mix} + {\rm subleading} \\
&\geq \min_{i} S_i + S_{\rm mix} + \mrm{subleading}\\
&=E_W(A:B) + \min_i \tfrac{1}{2}I(A':B')_i +S_{\rm mix}+ \mrm{subleading} \qquad \text{(from threads)} \\
&\geq E_W(A:B),
\end{aligned}\label{eq:lowerbound}
\ee
where $S_{\rm mix}$ is the entropy of mixing, $-\sum_i |\alpha_i|^2 \log|\alpha_i|^2$.
An important caveat  is that the superposition formula that we discuss above applies only when the number of terms $M$ in the superposition is small relative to $e^{O(c)}$. 
The necessity of this restriction can be understood from the fact that any product state can be written as a superposition of a large number of entangled geometric states, exploiting the overcompleteness of the basis, as in, e.g., \Ref{VanRaamsdonk:2010pw}. Therefore, an assumption must be made here to conclude the argument above: the number of terms in the superposition must still be small relative to when the approximation breaks down. 
An argument supporting this assumption follows from our reasoning for the existence of the decomposition in \Eq{eq:decomp} itself, using the construction involving black holes and subsequent row reduction of the basis states. That construction suggests that the number of terms in the superposition~\eqref{eq:decomp} will scale at most as $\dim \Hil_{(AB)^c}$. 
Because the total set of states for the entire boundary CFT scales as $\dim \Hil_{AB(AB)^c} \sim e^{O(c)}$, there will be a suppression in the entropy-of-mixing term relative to the leading term  that goes like
\be
\frac{S_{\rm mix}}{E_W(A:B)} \lesssim \frac{\log M}{c} \lesssim \frac{\log \dim \Hil_{(AB)^c}}{\log \dim \Hil_{AB(AB)^c}} < 1,\label{eq:inequality}
\ee
where the first inequality comes from $E_W\sim c$~\cite{Brown:1986nw} and the upper bound on the Shannon entropy, $S_\mrm{mix}\leq \log M$, the second inequality comes from the definition of the Hilbert space on $(AB)^c$ for CFTs, and the final inequality is definitionally true.
When the number of terms $M$ is small compared to $\dim \Hil_{(AB)^c}$, i.e., when $\ket{\psi}$ can be formed by the superposition of a small number of classical states, then the second inequality in \Eq{eq:inequality} can be strongly satisfied.
In \Eq{eq:inequality}, we are working in the formalism of a UV-regulated CFT, so that all Hilbert space dimensions are finite, making all the ratios well defined; holographically, this just corresponds to imposing a cutoff at fixed radial coordinate near the boundary, which also makes all of the geometric quantities we consider finite.

One potentially illuminating way of thinking about this is to consider the reason why the block-diagonal approximation of $\ket \psi$ is appropriate for sufficiently small superpositions. The reason is because the off-diagonal terms are suppressed by $e^c$ relative to the sum of the diagonal terms of $\ketbra{\psi}{\psi}$. This suppression only becomes weak enough to merit consideration of the off-diagonal terms when the number of nontrivial diagonal terms approaches $e^c$; notably, if the number is only $e^{kc}$ for $k<1$, there should still be a relative exponential suppression of the off-diagonal elements by $e^{(k-1)c}$. This suggests that the inequality in \Eq{eq:inequality} may in fact be even stronger than it appears, as it deals in terms of entropies as opposed to Hilbert~space dimensionality.

Putting together Eqs.~\eqref{eq:upperbound} and \eqref{eq:lowerbound}, we obtain
\be
E_P(A:B) = E_W(A:B). 
\ee
We have thus proven the $E_P = E_W$ conjecture, under the assumption that we can associate geometric purifications with bit thread configurations in the sense of \Sec{sec:upperbound} and, as motivated in this section, that our state $\ket{\psi}$ realizing the optimal purification can be written as a superposition of fewer than $\dim \Hil$ geometric states, each of which has fixed reduced density matrix on $AB$ corresponding to the entanglement wedge geometry.

\section{Discussion}\label{sec:discussion}

It should be made clear that some of the assumptions in \Sec{sec:lowerbound} have particularly strong support in three-dimensional gravity, where both a) the triviality of the Weyl tensor straightforwardly allows for the gluing of arbitrary black hole handles and b) the validity of Cardy's formula for density of states~\cite{Cardy:1986ie} is most compelling. We nevertheless expect some, probably more complicated, version of these arguments to persist in higher dimensions;\footnote{Indeed, as we have noted above, the gluing construction can be generalized using the formalism of Refs.~\cite{EW,Faulkner}.} we leave such investigations to future work. It would be a highly surprising development if subregion overcompleteness of geometric states were true only in three dimensions.

It is also worth stressing that the program advocated in this paper is not equivalent to the surface/state correspondence~\cite{miyaji2015surface}. Nowhere in the present work was it necessary to localize Hilbert space subfactors to the RT surface or indeed to any bulk surface; the entire argument was made from the perspective of boundary Hilbert space subfactors.
However, if we do assume the surface/state correspondence, then we do not need to extend the bit threads to an entire bulk geometry, since we can then localize the purifying degrees of freedom to the boundary of the entanglement wedge. In this case, the identification of the purification becomes trivial, with $A'$ (or $B'$) corresponding to the portions of $\gamma_{AB}$ between $A$ (or $B$, respectively) and $\Gamma$. The connection between the surface/state correspondence and the $E_P = E_W$ conjecture was first pointed out in \Ref{Takayanagi:2017knl}.

We also expect that the bit threads justification for $E_P=E_W$ will extend to the multipartite generalizations of entanglement of purification studied in Refs.~\cite{ bao2018holographic, bao2018conditional, bao2019entanglement, umemoto2018entanglement}. In particular, because we are classifying flows through surfaces, the same analysis should follow in these cases. In the multipartite generalization, the property that the minimal polytope in the bulk is inscribed within the RT surfaces has the consequence that, in the bit thread construction, all bit threads must cross the minimal polytope an even number of times (including zero), which allows for the RT surfaces to be completely partitioned into the $A_i'$. We leave investigation of the bit thread picture for multipartite entanglement of purification to future work.

\begin{center}
 {\bf Acknowledgments}
\end{center}
We thank Charles Cao, Illan Halpern, Matt Headrick, Yasunori Nomura, Nico Salzetta, and Mark Van Raamsdonk for useful discussions and comments.  
We are especially grateful to Jamie Sully for numerous fruitful discussions and initial collaboration.
N.B. is supported by the National Science Foundation under grant number 82248-13067-44-PHPXH,
by the Department of Energy under grant number DE-SC0019380, and by New York State
Urban Development Corporation Empire State Development contract no. AA289.
A.C.-D. is supported by the KU Leuven C1 grant ZKD1118 C16/16/005, the National Science Foundation of Belgium (FWO) grant G.001.12 Odysseus, and by the European Research Council grant no. ERC-2013-CoG 616732 HoloQosmos.
J.P. is supported in part by the Simons Foundation and in part by the Natural Sciences and Engineering Research Council of Canada.
G.N.R. is supported by the Miller Institute for Basic Research in Science at the University of California, Berkeley.

\appendix

\section{Holographic Purification of Minimal Dimension}\label{app:mindim}

Consider a purification $|\psi\rangle_{AA'BB'}$ of a state $\rho_{AB}$---in a holographic theory with a UV cutoff---
that realizes the infimum of $S(AA')$ over all purifications of $\rho_{AB}$.
In this Appendix we argue that $|\psi\rangle_{AA'BB'}$ can be compressed in
the dimensionality of ${\cal H}_{A'}\otimes{\cal H}_{B'}$ to a new
purification $|\phi\rangle_{AA'BB'}$ such that ${\rm dim}({\cal H}_{A'}\otimes{\cal H}_{B'})=e^{S(\rho_{AB})}$,
the minimal possible dimension of a purification of $\rho_{AB}$.

We begin by noting that a distillation of Bell pairs between $AA'$
and $BB'$ for $|\psi\rangle_{AA'BB'}$ is an LOCC procedure, so it
preserves $S(AA')$. If we view $A$, $A'$, $B$, and $B'$ as each
being some large tensor product over a number of qubits, where we
imagine that each qubit is either in a Bell state or unentangled (i.e., ignoring issues of multipartite entanglement), then there should be no pairs of qubits, one in $A'$ and one in $B'$, that
are in a Bell state. In particular, this is true because, were such
pairs of qubits to exist, one could decrease $S(AA')$ by simply excising
such pairs of qubits from $|\psi\rangle_{AA'BB'}$, in contradiction
with the hypothesis that $|\psi\rangle_{AA'BB'}$ realizes the infimum
of $S(AA')$. That is, we should have $I(A':B')=0$.
More generally, if there were multipartite entanglement among subfactors of $A$, $A'$, $B$, and $B'$, the purifying system could be re-factored to eliminate any contribution to $I(A':B')$.
For illustration, supposing that there were tripartite entanglement between some subfactors $a \subset A$, $a' \subset A'$, and $b' \subset B'$,  then redefining $A'$ to include $a'$ and $b'$ eliminates the possible contribution to $I(A':B')$, while also lowering $S(AA')$.
Again, if this were possible the initial partition in fact could not have been optimal.

We now perform the distillation procedure, which results in a tensor
product of three pure states:
\begin{equation}
|\psi\rangle_{AA'BB'}\in{\cal H}_{A}\otimes{\cal H}_{B}\otimes{\cal H}_{A'}\otimes{\cal H}_{B'}\stackrel{{\rm LOCC}}{\rightarrow}|\phi\rangle_{AA'BB'}=|\phi\rangle_{A_{0}A'_{0}}\otimes|\phi\rangle_{B_{0}B'_{0}}\otimes|\phi\rangle_{{\rm bell}},\label{eq:firstdecomp}
\end{equation}
where
\begin{equation}
\begin{aligned}{\cal H}_{A} & ={\cal H}_{A_{0}}\otimes{\cal H}_{A_{{\rm bell}}}\\
{\cal H}_{B} & ={\cal H}_{B_{0}}\otimes{\cal H}_{B_{{\rm bell}}}\\
{\cal H}_{A'} & ={\cal H}_{A'_{0}}\otimes{\cal H}_{A'_{{\rm bell}}}\\
{\cal H}_{B'} & ={\cal H}_{B'_{0}}\otimes{\cal H}_{B'_{{\rm bell}}}
\end{aligned}
\label{eq:factors}
\end{equation}
and where
\begin{equation}
\begin{aligned}|\phi\rangle_{{\rm bell}} & \in{\cal H}_{A_{{\rm bell}}}\otimes{\cal H}_{B_{{\rm bell}}}\otimes{\cal H}_{A'_{{\rm bell}}}\otimes{\cal H}_{B'_{{\rm bell}}}\\
|\phi\rangle_{A_{0}A'_{0}} & \in{\cal H}_{A_{0}}\otimes{\cal H}_{A'_{0}}\\
|\phi\rangle_{B_{0}B'_{0}} & \in{\cal H}_{B_{0}}\otimes{\cal H}_{B'_{0}}.
\end{aligned}
\end{equation}
Note that, at this point, none of the Hilbert space factors are ``prunable,''
in the sense that none of the factors of $|\phi\rangle_{AA'BB'}$
lives exclusively in ${\cal H}_{A_{0}'}$ or ${\cal H}_{B_{0}'}$
(and we recall that we are not allowed to delete any data about $\rho_{AB}$,
lest we jeopardize the recoverability of our original density matrix). 

\begin{figure}[ht]
\centering
\includegraphics[scale=0.5]{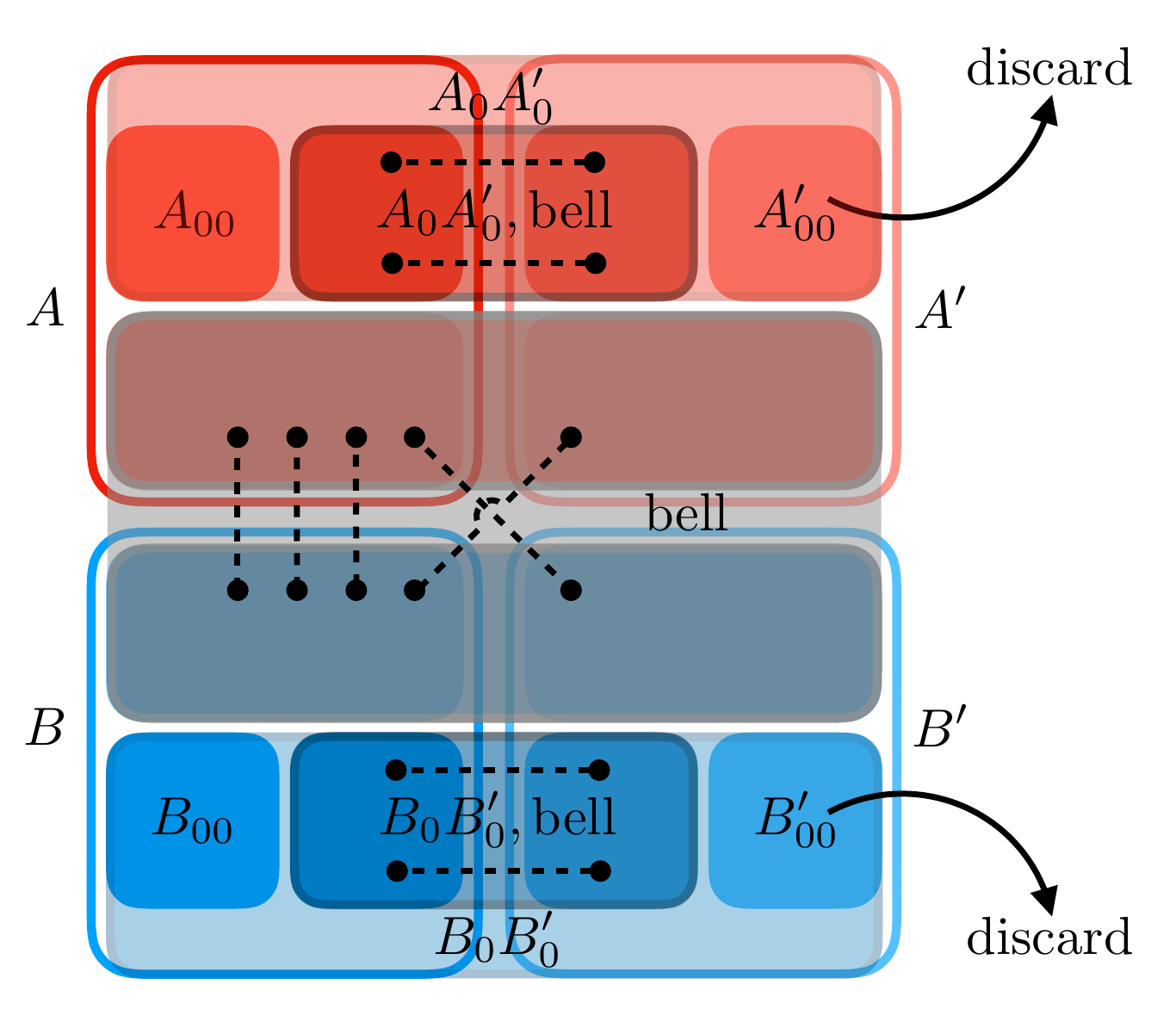}
\caption{Hilbert space decomposition and entanglement distillation of $AB$ and its purification.}
\label{fig:appendix}
\end{figure}

Next, let us distill twice more on $|\phi\rangle_{A_{0}A'_{0}}$ and
$|\phi\rangle_{B_{0}B'_{0}}$:
\begin{equation}
\begin{aligned}|\phi\rangle_{A_{0}A'_{0}} & \stackrel{{\rm LOCC}}{\rightarrow}|\phi\rangle_{A_{00}}\otimes|\phi\rangle_{A'_{00}}\otimes|\phi\rangle_{A_{0}A'_{0},{\rm bell}}\\
|\phi\rangle_{B_{0}B'_{0}} & \stackrel{{\rm LOCC}}{\rightarrow}|\phi\rangle_{B_{00}}\otimes|\phi\rangle_{B'_{00}}\otimes|\phi\rangle_{B_{0}B_{0}',{\rm bell}},
\end{aligned}
\label{eq:LOCC2}
\end{equation}
where
\begin{equation}
\begin{aligned}{\cal H}_{A_{0}} & ={\cal H}_{A_{00}}\otimes{\cal H}_{A_{0},{\rm bell}}\\
{\cal H}_{B_{0}} & ={\cal H}_{B_{00}}\otimes{\cal H}_{B_{0},{\rm bell}}\\
{\cal H}_{A'_{0}} & ={\cal H}_{A'_{00}}\otimes{\cal H}_{A'_{0},{\rm bell}}\\
{\cal H}_{B'_{0}} & ={\cal H}_{B'_{00}}\otimes{\cal H}_{B'_{0},{\rm bell}}
\end{aligned}
\end{equation}
and where
\begin{equation}
\begin{aligned}|\phi\rangle_{A_{0}A'_{0},{\rm bell}} & \in{\cal H}_{A_{0},{\rm bell}}\otimes{\cal H}_{A'_{0},{\rm bell}}\\
|\phi\rangle_{B_{0}B'_{0},{\rm bell}} & \in{\cal H}_{B_{0},{\rm bell}}\otimes{\cal H}_{B'_{0},{\rm bell}}\\
|\phi\rangle_{A_{00}} & \in{\cal H}_{A_{00}}\\
|\phi\rangle_{B_{00}} & \in{\cal H}_{B_{00}}\\
|\phi\rangle_{A'_{00}} & \in{\cal H}_{A'_{00}}\\
|\phi\rangle_{B'_{00}} & \in{\cal H}_{B'_{00}}.
\end{aligned}\label{eq:finaldecomp}
\end{equation}
Note that these distillations will also not affect $S(AA')$, since
the LOCCs in Eq.~\eqref{eq:LOCC2} are acting only on subfactors of
the Hilbert space that are unused in calculating $S(AA')$, as the
latter depends only on $|\phi\rangle_{{\rm bell}}$. See \Fig{fig:appendix} for an illustration of the Hilbert space decomposition and entanglement distillation in Eqs.~\eqref{eq:firstdecomp} through \eqref{eq:finaldecomp}.

Now, we are in a position to ``prune'' the Hilbert space factors
${\cal H}_{A_{00}'}$ and ${\cal H}_{B_{00}'}$, since these have
been rendered superfluous. Let us then enumerate the total remaining
Hilbert space dimensions of the ``primed'' factors. We note that
the remaining ``primed'' factors of the Hilbert space now each contain
a state that is maximally mixed and purified by some analogous unprimed
factor. This means that the Hilbert space dimension of the remaining
primed factors (i.e., less ${\cal H}_{A_{00}'}\otimes{\cal H}_{B_{00}'}$)
equals $e^{S(\rho_{AB})}$, because the state on the remaining primed
factors is simply a maximally-mixed purification of a state $\sigma_{AB}$
obtained by tracing out all the remaining primed factors from our
state we have after all the LOCC steps, for which $S(\sigma_{AB})=S(\rho_{AB})$.

We now note several additional facts. First, holographically, $\rho_{AB}$ is recoverable to leading order from
$\sigma_{AB}$ via a suitable unitary transformation, which follows from the
fact that $S(\sigma_{AB})=S(\rho_{AB})$ to leading order by the Hayden-Swingle-Walter
sandwiching result~\cite{HSW} (for a proof sketch, see Sec.\ 2.3 of \Ref{Bao:2018pvs}). It was also necessary that there existed no entanglement
between the reduced states on ${\cal H}_{A'_{{\rm bell}}}$ and ${\cal H}_{B'_{{\rm bell}}}$,
in order to guarantee the maximal mixing condition that led to our
conclusion of minimality of dimension. Having started with a state
$|\psi\rangle_{AA'BB'}$ with no mutual information between $A'$
and $B'$ (which we argued was a consequence of $|\psi\rangle_{AA'BB'}$
achieving the infimum of $S(AA')$), the Bell pair distillation gains
no advantage by introducing correlation between $A'_{{\rm bell}}$
and $B'_{{\rm bell}}$, so this condition should follow.

In addition, in order for $\rho_{AB}$ to be recoverable after our
LOCC steps, it is necessary for the entanglement cost of $\rho_{AB}$
to equal the distillable entanglement of $\rho_{AB}$. Luckily, this
equality was shown for holographic states (like $\rho_{AB}$) in \Ref{Bao:2018pvs}.

In summary, given a purification of the holographic state on $AB$, such that the purification minimizes
$S(AA')$, we can exhibit a purification, with the
same $S(AA')$, that also attains the minimal possible Hilbert space dimension.

\noindent

\bibliographystyle{utphys-modified}
\bibliography{EP-bit_threads}

\providecommand{\href}[2]{#2}\begingroup\raggedright\begin{thebibliography}{10}

\bibitem{tHooft:1993dmi}
G.~'t~Hooft, ``{Dimensional reduction in quantum gravity},'' in {\em
  {Conference on Highlights of Particle and Condensed Matter Physics
  (SALAMFEST)}}, vol.~C930308, p.~284.
\newblock 1993.
\newblock
\href{http://arxiv.org/abs/gr-qc/9310026}{{\ttfamily arXiv:gr-qc/9310026
  [gr-qc]}}.
\newblock

\bibitem{Susskind:1994vu}
L.~Susskind, ``{The world as a hologram},''
  \href{http://dx.doi.org/10.1063/1.531249}{{\em J. Math. Phys.} {\bfseries 36}
  (1995) 6377},
\href{http://arxiv.org/abs/hep-th/9409089}{{\ttfamily arXiv:hep-th/9409089
  [hep-th]}}.

\bibitem{Bousso:2002ju}
R.~Bousso, ``{The holographic principle},''
  \href{http://dx.doi.org/10.1103/RevModPhys.74.825}{{\em Rev. Mod. Phys.}
  {\bfseries 74} (2002) 825},
\href{http://arxiv.org/abs/hep-th/0203101}{{\ttfamily arXiv:hep-th/0203101
  [hep-th]}}.

\bibitem{Ryu:2006bv}
S.~Ryu and T.~Takayanagi, ``{Holographic derivation of entanglement entropy
  from AdS/CFT},'' \href{http://dx.doi.org/10.1103/PhysRevLett.96.181602}{{\em
  Phys. Rev. Lett.} {\bfseries 96} (2006) 181602},
\href{http://arxiv.org/abs/hep-th/0603001}{{\ttfamily arXiv:hep-th/0603001
  [hep-th]}}.

\bibitem{Ryu:2006ef}
S.~Ryu and T.~Takayanagi, ``{Aspects of holographic entanglement entropy},''
  \href{http://dx.doi.org/10.1088/1126-6708/2006/08/045}{{\em JHEP} {\bfseries
  08} (2006) 045},
\href{http://arxiv.org/abs/hep-th/0605073}{{\ttfamily arXiv:hep-th/0605073
  [hep-th]}}.

\bibitem{Lewkowycz:2013nqa}
A.~Lewkowycz and J.~Maldacena, ``{Generalized gravitational entropy},''
  \href{http://dx.doi.org/10.1007/JHEP08(2013)090}{{\em JHEP} {\bfseries 08}
  (2013) 090},
\href{http://arxiv.org/abs/1304.4926}{{\ttfamily arXiv:1304.4926 [hep-th]}}.

\bibitem{Maldacena:1997re}
J.~M. Maldacena, ``{The large-$N$ limit of superconformal field theories and
  supergravity},'' \href{http://dx.doi.org/10.1023/A:1026654312961}{{\em Int.
  J. Theor. Phys.} {\bfseries 38} (1999) 1113},
\href{http://arxiv.org/abs/hep-th/9711200}{{\ttfamily arXiv:hep-th/9711200
  [hep-th]}}.

\bibitem{Gubser:1998bc}
S.~S. Gubser, I.~R. Klebanov, and A.~M. Polyakov, ``{Gauge theory correlators
  from noncritical string theory},''
  \href{http://dx.doi.org/10.1016/S0370-2693(98)00377-3}{{\em Phys. Lett.}
  {\bfseries B428} (1998) 105},
\href{http://arxiv.org/abs/hep-th/9802109}{{\ttfamily arXiv:hep-th/9802109
  [hep-th]}}.

\bibitem{Witten:1998qj}
E.~Witten, ``{Anti-de Sitter space and holography},''
  \href{http://dx.doi.org/10.4310/ATMP.1998.v2.n2.a2}{{\em Adv. Theor. Math.
  Phys.} {\bfseries 2} (1998) 253},
\href{http://arxiv.org/abs/hep-th/9802150}{{\ttfamily arXiv:hep-th/9802150
  [hep-th]}}.

\bibitem{Aharony:1999ti}
O.~Aharony, S.~S. Gubser, J.~Maldacena, H.~Ooguri, and Y.~Oz, ``{Large $N$
  field theories, string theory and gravity},''
  \href{http://dx.doi.org/10.1016/S0370-1573(99)00083-6}{{\em Phys. Rept.}
  {\bfseries 323} (2000) 183},
\href{http://arxiv.org/abs/hep-th/9905111}{{\ttfamily arXiv:hep-th/9905111
  [hep-th]}}.

\bibitem{terhal2002entanglement}
B.~M. Terhal, M.~Horodecki, D.~W. Leung, and D.~P. DiVincenzo, ``The
  entanglement of purification,''
  \href{http://dx.doi.org/10.1063/1.1498001}{{\em J. Math. Phys.} {\bfseries
  43} (2002) 4286}, \href{http://arxiv.org/abs/quant-ph/0202044}{{\ttfamily
  arXiv:quant-ph/0202044 [quant-ph]}}.

\bibitem{Takayanagi:2017knl}
T.~Takayanagi and K.~Umemoto, ``{Entanglement of purification through
  holographic duality},''
  \href{http://dx.doi.org/10.1038/s41567-018-0075-2}{{\em Nature Phys.}
  {\bfseries 14} (2018) 573},
\href{http://arxiv.org/abs/1708.09393}{{\ttfamily arXiv:1708.09393 [hep-th]}}.

\bibitem{nguyen2018entanglement}
P.~Nguyen, T.~Devakul, M.~G. Halbasch, M.~P. Zaletel, and B.~Swingle,
  ``{Entanglement of purification: from spin chains to holography},''
  \href{http://dx.doi.org/10.1007/JHEP01(2018)098}{{\em JHEP} {\bfseries 01}
  (2018) 098},
\href{http://arxiv.org/abs/1709.07424}{{\ttfamily arXiv:1709.07424 [hep-th]}}.

\bibitem{Agon:2018lwq}
C.~A. Ag\'{o}n, J.~De~Boer, and J.~F. Pedraza, ``{Geometric Aspects of
  Holographic Bit Threads},''
\href{http://arxiv.org/abs/1811.08879}{{\ttfamily arXiv:1811.08879 [hep-th]}}.

\bibitem{bao2018holographic}
N.~Bao and I.~F. Halpern, ``{Holographic Inequalities and Entanglement of
  Purification},'' \href{http://dx.doi.org/10.1007/JHEP03(2018)006}{{\em JHEP}
  {\bfseries 03} (2018) 006},
\href{http://arxiv.org/abs/1710.07643}{{\ttfamily arXiv:1710.07643 [hep-th]}}.

\bibitem{bao2018conditional}
N.~Bao and I.~F. Halpern, ``{Conditional and Multipartite Entanglements of
  Purification and Holography},''
  \href{http://dx.doi.org/10.1103/PhysRevD.99.046010}{{\em Phys. Rev.}
  {\bfseries D99} (2019) 046010},
\href{http://arxiv.org/abs/1805.00476}{{\ttfamily arXiv:1805.00476 [hep-th]}}.

\bibitem{bao2019entanglement}
N.~Bao, A.~Chatwin-Davies, and G.~N. Remmen, ``{Entanglement of Purification
  and Multiboundary Wormhole Geometries},''
  \href{http://dx.doi.org/10.1007/JHEP02(2019)110}{{\em JHEP} {\bfseries 02}
  (2019) 110},
\href{http://arxiv.org/abs/1811.01983}{{\ttfamily arXiv:1811.01983 [hep-th]}}.

\bibitem{umemoto2018entanglement}
K.~Umemoto and Y.~Zhou, ``{Entanglement of Purification for Multipartite States
  and its Holographic Dual},''
  \href{http://dx.doi.org/10.1007/JHEP10(2018)152}{{\em JHEP} {\bfseries 10}
  (2018) 152},
\href{http://arxiv.org/abs/1805.02625}{{\ttfamily arXiv:1805.02625 [hep-th]}}.

\bibitem{Faulkner}
S.~Dutta and T.~Faulkner, ``{A canonical purification for the entanglement
  wedge cross-section},''
\href{http://arxiv.org/abs/1905.00577}{{\ttfamily arXiv:1905.00577 [hep-th]}}.

\bibitem{Kudler-Flam:2019oru}
J.~Kudler-Flam, I.~MacCormack, and S.~Ryu, ``{Holographic entanglement contour,
  bit threads, and the entanglement tsunami},''
\href{http://arxiv.org/abs/1902.04654}{{\ttfamily arXiv:1902.04654 [hep-th]}}.

\bibitem{Du:2019emy}
D.-H. Du, C.-B. Chen, and F.-W. Shu, ``{Bit threads and holographic
  entanglement of purification},''
\href{http://arxiv.org/abs/1904.06871}{{\ttfamily arXiv:1904.06871 [hep-th]}}.

\bibitem{Jokela:2019ebz}
N.~Jokela and A.~P{\" o}nni, ``{Notes on entanglement wedge cross sections},''
\href{http://arxiv.org/abs/1904.09582}{{\ttfamily arXiv:1904.09582 [hep-th]}}.

\bibitem{Freedman:2016zud}
M.~Freedman and M.~Headrick, ``{Bit threads and holographic entanglement},''
  \href{http://dx.doi.org/10.1007/s00220-016-2796-3}{{\em Commun. Math. Phys.}
  {\bfseries 352} (2017) 407},
\href{http://arxiv.org/abs/1604.00354}{{\ttfamily arXiv:1604.00354 [hep-th]}}.

\bibitem{Cui:2018dyq}
S.~X. Cui, P.~Hayden, T.~He, M.~Headrick, B.~Stoica, and M.~Walter, ``{Bit
  Threads and Holographic Monogamy},''
\href{http://arxiv.org/abs/1808.05234}{{\ttfamily arXiv:1808.05234 [hep-th]}}.

\bibitem{Harlow:2016vwg}
D.~Harlow, ``{The Ryu-Takayanagi Formula from Quantum Error Correction},''
  \href{http://dx.doi.org/10.1007/s00220-017-2904-z}{{\em Commun. Math. Phys.}
  {\bfseries 354} (2017) 865},
\href{http://arxiv.org/abs/1607.03901}{{\ttfamily arXiv:1607.03901 [hep-th]}}.

\bibitem{Bakhmatov:2017ihw}
I.~Bakhmatov, N.~S. Deger, J.~Gutowski, E.~$\text{\'{O}\;Colg\'{a}in}$, and
  H.~Yavartanoo, ``{Calibrated Entanglement Entropy},''
  \href{http://dx.doi.org/10.1007/JHEP07(2017)117}{{\em JHEP} {\bfseries 07}
  (2017) 117},
\href{http://arxiv.org/abs/1705.08319}{{\ttfamily arXiv:1705.08319 [hep-th]}}.

\bibitem{Harvey:1982xk}
R.~Harvey and H.~B. Lawson, Jr., ``{Calibrated geometries},''
\href{http://dx.doi.org/10.1007/BF02392726}{{\em Acta Math.} {\bfseries 148}
  (1982) 47}.

\bibitem{Federer}
H.~Federer, ``{Real Flat Chains, Cochains and Variational Problems},''
  \href{http://dx.doi.org/10.1512/iumj.1975.24.24031}{{\em Indiana Univ. Math.
  J.} {\bfseries 24} (1975) 351}.

\bibitem{Strang1983}
G.~Strang, ``Maximal flow through a domain,''
  \href{http://dx.doi.org/10.1007/BF02592050}{{\em Mathematical Programming}
  {\bfseries 26} (1983) 123}.

\bibitem{MR1088184}
R.~Nozawa, ``Max-flow min-cut theorem in an anisotropic network,''
  \href{http://dx.doi.org/10.18910/7180}{{\em Osaka J. Math.} {\bfseries 27}
  (1990) 805}.

\bibitem{EW}
N.~Engelhardt and A.~C. Wall, ``{Coarse Graining Holographic Black Holes},''
\href{http://arxiv.org/abs/1806.01281}{{\ttfamily arXiv:1806.01281 [hep-th]}}.

\bibitem{Fannes1973}
M.~Fannes, ``A continuity property of the entropy density for spin lattice
  systems,'' \href{http://dx.doi.org/10.1007/BF01646490}{{\em Communications in
  Mathematical Physics} {\bfseries 31} (1973) 291}.

\bibitem{Bao:2018pvs}
N.~Bao, G.~Penington, J.~Sorce, and A.~C. Wall, ``{Beyond Toy Models:
  Distilling Tensor Networks in Full AdS/CFT},''
\href{http://arxiv.org/abs/1812.01171}{{\ttfamily arXiv:1812.01171 [hep-th]}}.

\bibitem{HSW}
P.~Hayden, B.~Swingle, and M.~Walter, {\em forthcoming}.

\bibitem{miyaji2015surface}
M.~Miyaji and T.~Takayanagi, ``{Surface/State Correspondence as a Generalized
  Holography},'' \href{http://dx.doi.org/10.1093/ptep/ptv089}{{\em PTEP}
  {\bfseries 2015} (2015) 073B03},
\href{http://arxiv.org/abs/1503.03542}{{\ttfamily arXiv:1503.03542 [hep-th]}}.

\bibitem{Botta-Cantcheff:2017gys}
M.~Botta-Cantcheff and P.~J. Mart\'{i}nez, ``{Which quantum states are dual to
  classical spacetimes?},''
\href{http://arxiv.org/abs/1703.03483}{{\ttfamily arXiv:1703.03483 [hep-th]}}.

\bibitem{Garfinkle:1990eq}
D.~Garfinkle and A.~Strominger, ``{Semiclassical Wheeler wormhole
  production},''
\href{http://dx.doi.org/10.1016/0370-2693(91)90665-D}{{\em Phys. Lett.}
  {\bfseries B256} (1991) 146}.

\bibitem{Maldacena:2013xja}
J.~Maldacena and L.~Susskind, ``{Cool horizons for entangled black holes},''
  \href{http://dx.doi.org/10.1002/prop.201300020}{{\em Fortsch. Phys.}
  {\bfseries 61} (2013) 781},
\href{http://arxiv.org/abs/1306.0533}{{\ttfamily arXiv:1306.0533 [hep-th]}}.

\bibitem{Bao:2015nqa}
N.~Bao, J.~Pollack, and G.~N. Remmen, ``{Splitting Spacetime and Cloning
  Qubits: Linking No-Go Theorems across the ER=EPR Duality},''
  \href{http://dx.doi.org/10.1002/prop.201500053}{{\em Fortsch. Phys.}
  {\bfseries 63} (2015) 705},
\href{http://arxiv.org/abs/1506.08203}{{\ttfamily arXiv:1506.08203 [hep-th]}}.

\bibitem{Giddings:2009gj}
S.~B. Giddings and R.~A. Porto, ``{The Gravitational S-matrix},''
  \href{http://dx.doi.org/10.1103/PhysRevD.81.025002}{{\em Phys. Rev.}
  {\bfseries D81} (2010) 025002},
\href{http://arxiv.org/abs/0908.0004}{{\ttfamily arXiv:0908.0004 [hep-th]}}.

\bibitem{Witten:1999xp}
E.~Witten and S.-T. Yau, ``{Connectedness of the boundary in the AdS/CFT
  correspondence},'' \href{http://dx.doi.org/10.4310/ATMP.1999.v3.n6.a1}{{\em
  Adv. Theor. Math. Phys.} {\bfseries 3} (1999) 1635},
\href{http://arxiv.org/abs/hep-th/9910245}{{\ttfamily arXiv:hep-th/9910245
  [hep-th]}}.

\bibitem{Cardy:1986ie}
J.~L. Cardy, ``{Operator Content of Two-Dimensional Conformally Invariant
  Theories},''
\href{http://dx.doi.org/10.1016/0550-3213(86)90552-3}{{\em Nucl. Phys.}
  {\bfseries B270} (1986) 186}.

\bibitem{Maldacena:2001kr}
J.~M. Maldacena, ``{Eternal black holes in anti-de Sitter},''
  \href{http://dx.doi.org/10.1088/1126-6708/2003/04/021}{{\em JHEP} {\bfseries
  04} (2003) 021},
\href{http://arxiv.org/abs/hep-th/0106112}{{\ttfamily arXiv:hep-th/0106112
  [hep-th]}}.

\bibitem{Srednicki}
M.~Srednicki, ``{Chaos and quantum thermalization},''
  \href{http://dx.doi.org/10.1103/PhysRevE.50.888}{{\em Phys. Rev.} {\bfseries
  E50} (1994) 888}, \href{http://arxiv.org/abs/cond-mat/9403051}{{\ttfamily
  arXiv:cond-mat/9403051 [cond-mat]}}.

\bibitem{Almheiri:2016blp}
A.~Almheiri, X.~Dong, and B.~Swingle, ``{Linearity of Holographic Entanglement
  Entropy},'' \href{http://dx.doi.org/10.1007/JHEP02(2017)074}{{\em JHEP}
  {\bfseries 02} (2017) 074},
\href{http://arxiv.org/abs/1606.04537}{{\ttfamily arXiv:1606.04537 [hep-th]}}.

\bibitem{VanRaamsdonk:2010pw}
M.~Van~Raamsdonk, ``{Building up spacetime with quantum entanglement},''
  \href{http://dx.doi.org/10.1007/s10714-010-1034-0,
  10.1142/S0218271810018529}{{\em Gen. Rel. Grav.} {\bfseries 42} (2010) 2323},
  \href{http://arxiv.org/abs/1005.3035}{{\ttfamily arXiv:1005.3035 [hep-th]}}.
[Int. J. Mod. Phys.D19,2429(2010)].

\bibitem{Brown:1986nw}
J.~D. Brown and M.~Henneaux, ``{Central Charges in the Canonical Realization of
  Asymptotic Symmetries: An Example from Three-Dimensional Gravity},''
\href{http://dx.doi.org/10.1007/BF01211590}{{\em Commun. Math. Phys.}
  {\bfseries 104} (1986) 207}.

\end{thebibliography}\endgroup

\end{document}